\documentclass{WileyChemistry-template}
\usepackage{amsmath}

\title{\raggedright Weak Polar Optical Phonon Scattering Decouples Electron and Phonon Transport in Layered Thermoelectric Materials}

\author{
	\begin{minipage}{\textwidth}	
		Zhonghao Xia, Michele Reticcioli, Yateng Wang, Yali Yang, Alessandro Stroppa, and Jiangang He*
	\end{minipage}
}

\newcommand{\affiliation}{
	\begin{itemize}		
		\item[{[a]}] Z. Xia, Y. Wang, Prof. Y. Yang, Prof. J. He\\
		State Key Laboratory for Advanced Metals and Materials, Beijing Key Laboratory for Magneto-Photoelectrical Composite and Interface Science, School of Mathematics and Physics, University of Science and Technology Beijing, Beijing 100083, China\\
		E-mail: \textcolor{blue}{jghe2021@ustb.edu.cn}
		
		\item[{[b]}] M. Reticcioli, Prof. A. Stroppa
		CNR-SPIN c/o Department of Physical and Chemical Sciences, University of L’Aquila, Via Vetoio, I-67100, Coppito, L’Aquila, Italy
	\end{itemize}
}

\renewcommand{\abstract}{High-performance thermoelectric (TE) materials are crucial for efficient waste-heat recovery and solid-state cooling technologies. A persistent challenge in TE materials design arises from the strong interdependence among the electrical conductivity ($\sigma$), Seebeck coefficient ($S$), and lattice thermal conductivity ($\kappa_{\mathrm{L}}$). Layered compounds can effectively suppress $\kappa_{\mathrm{L}}$ along the cross-plane direction owing to weak interlayer interactions; however, they often suffer from low carrier mobility ($\mu$) caused by limited band dispersion and strong polar optical phonon (POP) scattering. Here, we perform high-throughput density functional theory calculations to screen 236 layered semiconductors and identify candidates with low effective mass ($m^{*}$) and weak POP scattering. We identify 23 compounds with high cross-plane $\mu$, among which 14 exhibit large power factors ($S^{2}\sigma$). Notably, GaGe$_{2}$Te stands out with exceptionally high cross-plane $\sigma$ and power factor, enabled by a favorable combination of small $m^{*}$ and a small ionic dielectric constant. Simultaneously, GaGe$_{2}$Te exhibits an ultralow cross-plane $\kappa_{\mathrm{L}}$ of 0.57~W~m$^{-1}$~K$^{-1}$ at 300~K, originating from weak interlayer bonding and pronounced phonon anharmonicity. These results demonstrate an effective strategy to decouple electron and phonon transport in layered materials by mitigating POP scattering, thereby providing a promising pathway toward high-performance thermoelectric materials.}

\begin{document}
	
	\begin{sloppypar}
	%%%%%%%%%%%%%%%%%%%%%%%%%%%%%%%%%%%%%%%%%%%%%%%%%%%%%%%%%%
	%%%%%%%%%%%%%%%%%%%%%%%%%%%%%%%%%%%%%%%%%%%%%%%%%%%%%%%%%%
	%%%%%%%%%%%%%%%%%%%%%%%%%%%%%%%%%%%%%%%%%%%%%%%%%%%%%%%%%%
	
	\twocolumn[\vspace{-1.5cm}\maketitle\vspace{-1cm}
	\textit{\dedication}\vspace{0.4cm}]
	\small{\begin{shaded}
			\noindent\abstract
	\end{shaded}}

	\begin{figure} [!b]
		\begin{minipage}[t]{\columnwidth}{\rule{\columnwidth}{1pt}\footnotesize{\textsf{\affiliation}}}\end{minipage}
	\end{figure}
	
\section*{Introduction}\label{Introduction}
	Thermoelectric technology, capable of directly and reversibly converting heat into electricity, offers a promising solution for waste heat recovery, solid-state cooling, and precise temperature control~\cite{burnete2022review,qin2022solid,zheng2022durable,crane2004optimization}. However, the relatively low energy conversion efficiency of current TE materials has severely limited their widespread practical deployment. Consequently, the discovery and rational design of novel TE materials with enhanced performance have emerged as central objectives in the field.
	
	The efficiency of TE materials is quantified by the dimensionless figure of merit, $\mathrm{ZT} = S^2\sigma \mathrm{T}/(\kappa_\mathrm{e} + \kappa_\mathrm{L})$, where $S$ is the Seebeck coefficient, $\sigma$ is the electrical conductivity, $\mathrm{T}$ is the absolute temperature, and $\kappa_\mathrm{e}$ and $\kappa_\mathrm{L}$ denote the electronic and lattice thermal conductivities, respectively. Enhancing $\mathrm{ZT}$ requires increasing the power factor ($\mathrm{PF} = S^2\sigma$) while simultaneously reducing $\kappa_\mathrm{L}$. However, due to the intrinsic interdependence among these transport parameters, achieving substantial improvements in $\mathrm{ZT}$ remains a formidable challenge in both fundamental research and practical material design. 
	In semiconductors, the electrical conductivity $\sigma$ is proportional to the carrier mobility ($\mu$) and the relaxation time ($\tau$), and inversely proportional to the inertial effective mass ($m_{\mathrm{I}}^*$): $\sigma = ne\mu = \frac{ne^2\tau}{m_{\mathrm{I}}^*}$, where $n$ denotes the carrier concentration and $e$ is the elementary charge. Accordingly, achieving a high $\sigma$ generally favors a small $m_{\mathrm{I}}^*$. In contrast, a large Seebeck coefficient $S$ typically benefits from a large density-of-states effective mass ($m_{\mathrm{d}}^*$)~\cite{2018A}. Although a high valley degeneracy ($N_{\mathrm{v}}$) can, in principle, mitigate this trade-off---since $m_{\mathrm{d}}^* \propto N_{\mathrm{v}}^{2/3}$ while the total conductivity scales approximately linearly with $N_{\mathrm{v}}$---for systems with limited $N_{\mathrm{v}}$, the competing requirements of $S$ and $\sigma$ imposed by the electronic structure remain a primary bottleneck to achieving high-$\mathrm{ZT}$ performance.
	Furthermore, semiconductors with low $\kappa_{\mathrm{L}}$ often exhibit low $\sigma$, because $\sigma$ scales with elastic stiffness (e.g., the bulk modulus) within the acoustic deformation potential scattering regime~\cite{PhysRev.80.72}, whereas a large bulk modulus typically reflects strong bonding interactions that lead to high $\kappa_{\mathrm{L}}$~\cite{https://doi.org/10.1002/anie.201508381}.
	
	Numerous strategies have been employed to suppress $\kappa_{\mathrm{L}}$, including defect engineering~\cite{Mao03042018,2018Nano,2008High}, nanoscale precipitates~\cite{doi:10.1126/science.1092963,2012High,2008Enhanced,singh2024enhanced}, lone-pair electrons~\cite{PhysRevLett.107.235901,2013Lone}, rattling phonon modes~\cite{2015Impact,2016Ultralow,2021Physical}, bond heterogeneity~\cite{powell2025exploiting}, resonant four-phonon scattering~\cite{PhysRevLett.125.245901}, and weak chemical bonding~\cite{https://doi.org/10.1002/adfm.202108532,https://doi.org/10.1002/advs.202417292,he2019designing}. Meanwhile, band convergence and high valley degeneracy have been exploited to enhance PF~\cite{2012Convergence,moshwan2019realizing,tan2017improving,long2023band,pei2012thermoelectric,doi:10.1021/jacs.4c04048,kim2017high,he2019designing,xiong2025forbidden}. Nevertheless, only a limited number of approaches can optimize PF and $\kappa_{\mathrm{L}}$ concurrently, embodying the ``phonon-glass electron-crystal'' (PGEC) paradigm~\cite{nolas1999skutterudites}.
	
	Layered and quasi-one-dimensional (1D) materials often exhibit pronounced anisotropy in both electronic and phononic transport, providing an alternative route to decouple $\sigma$ and $\kappa_\mathrm{L}$. Beyond weak vdW interlayer interactions, these anisotropic structures intrinsically incorporate effective phonon-engineering mechanisms, such as bond-heterogeneity-induced anharmonicity and phonon band flattening. Their weak interlayer or interchain interactions and inherent structural anisotropy can significantly reduce phonon group velocities ($v_{\mathrm{g}}$), thereby effectively suppressing $\kappa_\mathrm{L}$. If a high $\sigma$ can be achieved along the cross-plane or cross-chain direction, substantial improvements in $\mathrm{ZT}$ become attainable, although many layered compounds exhibit relatively small cross-plane band dispersion. Previous studies shown that SnSe~\cite{doi:10.1126/science.aaq1479}, Mg$_3$Sb$_2$~\cite{li2021demonstration}, Na$_2$Au$X$ ($X$ = P, As, Sb, and Bi)~\cite{xia2025strong} exhibit high PF along the cross-plane (cross-chain) direction, primarily due to their large band dispersion (small effective mass) and multiband electronic structures with high valley degeneracy along the cross-plane (cross-chain) direction.
	
	In addition to effective mass, the relaxation time $\tau$ plays a critical role in carrier transport. Although strategies to increase $\tau$ have been relatively underexplored, they are of significant importance for high-performance TE materials, as longer $\tau$ directly translates to higher $\sigma$. Electron--phonon interactions constitute the dominant intrinsic mechanism governing carrier transport in solids. From a computational perspective, electron--phonon coupling generally comprises short-range deformation-potential interactions and long-range polar optical phonon coupling mediated by the Fr\"{o}hlich interaction~\cite{park2025advances}. In polar semiconductors, the macroscopic electric fields generated by longitudinal optical phonons interact strongly with charge carriers, making polar optical phonon (POP) scattering the dominant mobility-limiting mechanism at elevated temperatures.
	%	From a computational perspective, the electron--phonon coupling matrix can be decomposed into two primary contributions~\cite{park2025advances}: $g^{\mathrm{e-ph}} = g^{\mathrm{DP}} + g^{\mathrm{PO}}$, where $g^{\mathrm{DP}}$ represents deformation-potential (DP) scattering arising from atomic vibrations via short-range interactions, and $g^{\mathrm{PO}}$ describes polar optical phonon (POP) scattering originating from internal electric fields induced by atomic displacements in polar materials. In polar crystals, distinct cation and anion sublattices give rise to strongly polarized optical phonon modes. Upon excitation of longitudinal optical (LO) phonons, oppositely charged ions oscillate out of phase along the phonon propagation direction, generating a macroscopic polarization field. This built-in electric field couples strongly with free carriers through long-range Coulomb interactions, significantly reducing $\mu$, particularly at elevated temperatures. 
	Therefore, layered materials featuring small effective mass or multiple energy valleys along the cross-plane direction, combined with intrinsically weak POP scattering to enhance carrier relaxation time, represent promising candidates for high-performance thermoelectrics.
	
	Herein, we establish a quantitative framework linking chemical bonding character to $\mu$ in layered semiconductors through high-throughput first-principles calculations. By screening 236 thermodynamically stable semiconductors across 22 structural prototypes and systematically analyzing their effective mass, dielectric responses, and crystal orbital bond indices (COBI), we identify 23 compounds with high cross-plane $\mu$, and 14 of them display large PF. Among them, GaGe$_2$Te exhibits an exceptional high cross-plane mobility ($\mu$ = 645~cm$^2$V$^{-1}$s$^{-1}$ at 300 K and hole concentration of 10$^{19}$ cm$^{-3}$), when all relevant scattering mechanisms, including acoustic deformation potential (ADP), POP, and ionized impurity (IMP) scattering, are considered. Our findings reveal that minimizing ionic dielectric constants and effective mass along the cross-plane direction can effectively enhance PF, thereby establishing ``bonding character engineering'' as a powerful design principle for next-generation high-performance TE materials.

	\section*{Computational Methods}\label{results_discussion}
	The density functional theory (DFT) calculations of lattice dynamics in this work were performed using the Vienna \textit{Ab initio} Simulation Package (VASP)~\cite{vasp1,vasp2}, employing the Projector Augmented-Wave (PAW) method~\cite{PAW1,PAW2}. The exchange–correlation interactions were treated using the PBEsol functional within the framework of the Generalized Gradient Approximation (GGA)~\cite{pbesol1}, which was found to be accurate enough to predict interlayer distance of layered compounds~\cite{PhysRevB.89.075409}. The cut-off energy of plane wave is set to 520 eV. A $\Gamma$-centered 12 $\times$ 12 $\times$ 6 k-point grid is used to sample the Brillouin zone. In addition, the convergence criteria for the total energy and the force acting on each atom are set to 10$^{-8}$ eV and 0.01 eV/\AA, respectively. To obtain the accurate band structure, the Heyd-Scuseria-Ernzerhof (HSE06) hybrid functional is used to remedy the exchange–correlation interaction\cite{10.1063/1.1760074}.
	The crystal orbital Hamilton population (COHP) and crystal orbital bond index (COBI) analysis were performed for selected nearest-neighbor atomic pairs using the LOBSTER code,~\cite{nelson2020lobster} based on the wavefunctions generated with the PBEsol exchange-correlation functional.
	The bulk modulus ($B$), shear moduli ($G$), and average speed of sound ($v_a$) were calculated from the elastic constants obtained through finite differences methods. The ion dielectric tensor and the Born effective charges are calculated using the density functional perturbation theory (DFPT). In order to get more accurate results, we obtain the high-frequency dielectric constant by calculating the self-consistent response under a finite electric field using the HSE06 functional. To more accurately characterize the aspects of the band structure most relevant to $\sigma$, we adopted the conductivity mass ($m_{\mathrm{c}}^*$) proposed by Gibbs \emph{et al.}~\cite{gibbs2017effective}. This quantity is evaluated from the electrical conductivity divided by relaxation time ($\sigma/\tau$), computed using BoltzTraP2~\cite{madsen2018boltztrap2} within the constant relaxation time approximation. The value of $m_{\mathrm{c}}^*$ was obtained by inverting the relation $\sigma/\tau = ne^2/m_{\mathrm{c}}^*$ at 300~K and a carrier concentration of $1 \times 10^{19}$~cm$^{-3}$.
	
	The second-order force constants were calculated using the finite displacement method as implemented in the Phonopy code~\cite{TOGO20151}. A 4 $\times$ 4 $\times$ 2 supercell and a 2 $\times$ 2 $\times$ 1 $k$-point mesh were utilized, with a displacement of 0.01 \AA. The random configurations used to extract the anharmonic force constants were generated from 20,000 steps of molecular dynamics (MD) simulations using force fields constructed through on-the-fly machine-learning training based on DFT calculations,~\cite{PhysRevB.100.014105} with a time step of 1 fs at 300 K. Then, 20 structures were selected from MD for high-precision DFT self-consistent calculations with $k$-point grid of 2 $\times$ 2 $\times$ 1. The compressive sensing lattice dynamics (CSLD)~\cite{RN38} were employed to extract third- and fourth-order force constants. The anharmonically renormalized phonon frequencies at finite temperatures were calculated using self-consistent phonon (SCPH) theory~\cite{werthamer1970self}. The thermal transport properties were obtained by solving the Peierls-Boltzmann transport equation through an iterative scheme as implemented in the FourPhonon package~\cite{HAN2022108179}. To ensure convergence, uniform 20 $\times$ 20 $\times$ 10 $q$-point meshes were employed to calculate the contributions from SCPH three-phonon scattering $\kappa_{\mathrm{3ph}}^{\mathrm{P}}$ and four-phonon scattering $\kappa_{\mathrm{3,4ph}}^{\mathrm{P}}$. Additionally, the calculation of the four-phonon scattering processes was accelerated using the sampling method~\cite{guo2024sampling}. The off-diagonal contributions from three-phonon scattering $\kappa_{\mathrm{3ph}}^{\mathrm{C}}$ and four-phonon scattering $\kappa_{\mathrm{ 3,4ph}}^{\mathrm{C}}$ were computed following the formalism developed by Simoncelli et al.~\cite{simoncelli2019unified}.
	
	The electron transport properties were calculated using the electron-phonon coupling theory~\cite{ziman2001electrons}. The AMSET software package was employed during high-throughput screening\cite{ganose2021efficient}, accounting for both ADP, POP and IMP scattering. For GaGe$_2$Te and the reference compound Mg$_3$Sb$_2$, DFT calculations using the Quantum Espresso (QE) Package\cite{qe}. The Optimized Norm-Conserving Vanderbilt (ONCV) scalar-relativistic pseudopotentials\cite{hamann2013optimized} with the PBEsol exchange-correlation functional are employed. The kinetic energy cutoff of 100 Ry and a 12$\times$12$\times$6 $k$-grid used to self-consistent and non-self-consistent calculations. The phonon dispersion is calculated using DFPT with 4$\times$4$\times$2 $q$-grid. After obtaining the electron and phonon eigenvalues, the electronic scattering rates can be calculated by combining the maximally localized Wannier functions (MLWF), obtained from the Wannier90 code~\cite{MOSTOFI20142309}, with the electron-phonon coupling, which are implemented in the Perturbo Code~\cite{zhou2021perturbo}. To achieve a converged scattering rate, a dense mesh of 120$\times$120$\times$60 $k$/$q$ points was used. In addition to the intrinsic phonon-limited scattering mechanisms, IMP was also included in the final transport calculations to account for extrinsic carrier scattering effects at finite carrier concentrations.

	\begin{figure}[tph!]
		\includegraphics[width=1.0\linewidth]{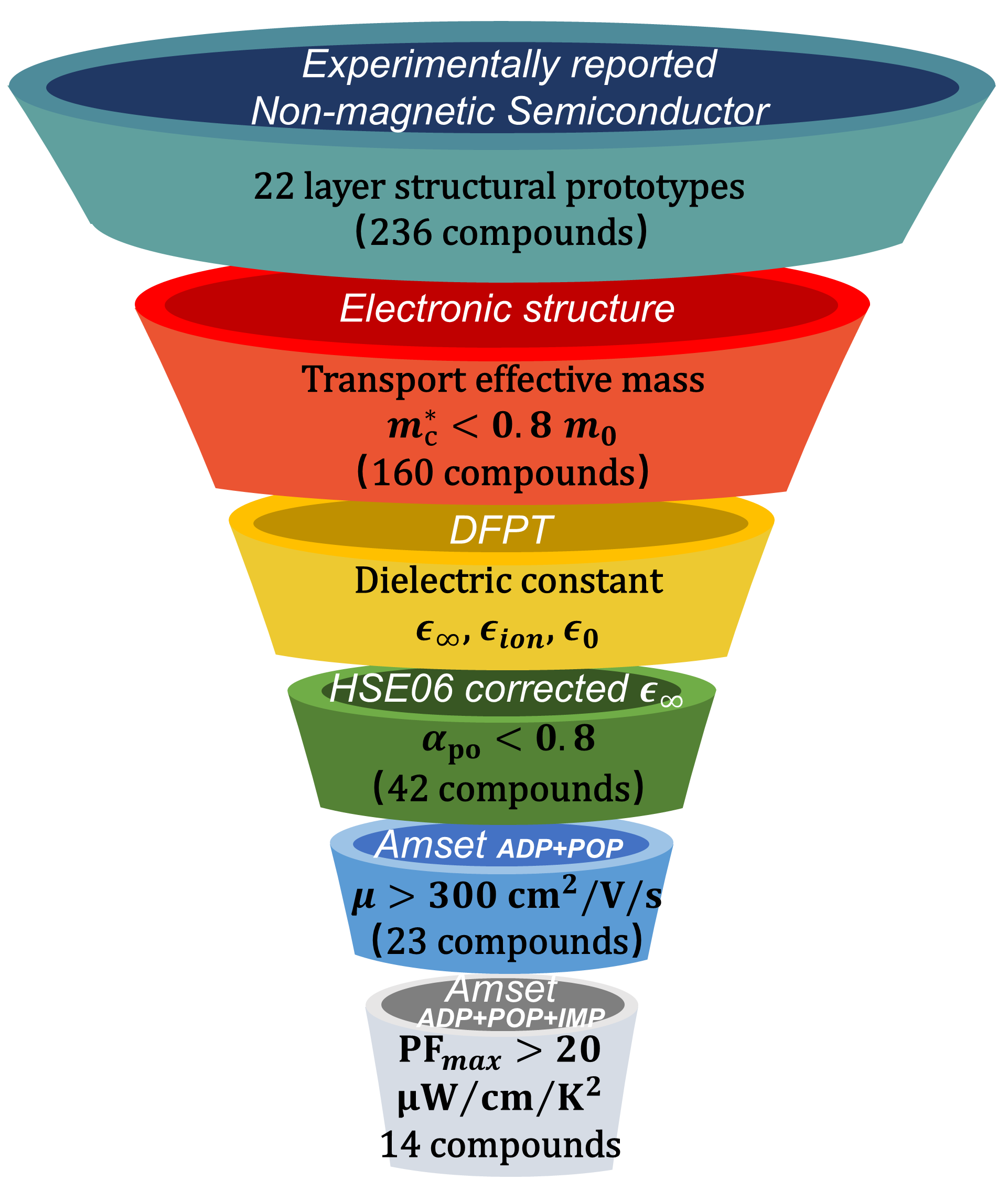}
		\caption{Screening workflow for identifying layered semiconductors with high carrier mobility and power factor at room temperature.}
		\label{fig1}
	\end{figure}
	
	\begin{figure*}[tph!]
		\includegraphics[width=0.8\linewidth]{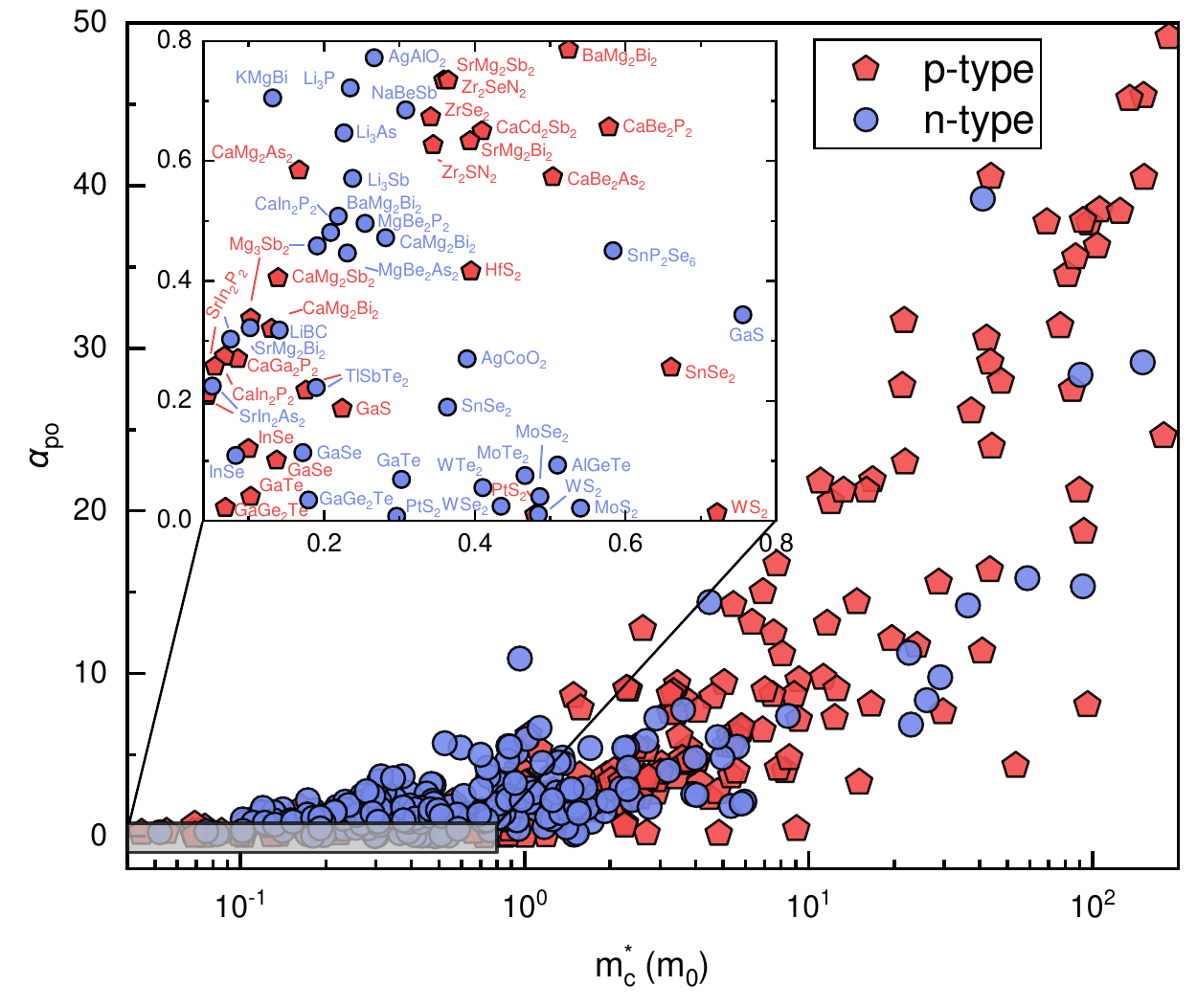}
		\caption{Distribution of the 236 screened compounds in the parameter space defined by the polar coupling constant $\alpha_{\mathrm{po}}$ and conductivity mass $m_{\mathrm{c}}^*$ along the cross-plane direction.}
		\label{fig2}
	\end{figure*}

	\section*{Results and Discussion}\label{results_discussion}
	\noindent \textbf{Material design strategy through suppressing Fr\"{o}hlich interaction.} The rational design of high-performance TE materials requires identifying intrinsic structural and electronic features that simultaneously enable high $\sigma$ while suppressing $\kappa_{\mathrm{L}}$. In polar crystals, the long-wavelength displacement of ions associated with longitudinal optical (LO) phonons induces a spatially oscillating electric field. This long-range Coulomb field couples to electronic states via the Fr\"{o}hlich interaction, producing an electron–phonon matrix element of the form~\cite{verdi2015frohlich}:
	\begin{equation}
		g_{\mathbf{q}} = \frac{i}{|\mathbf{q}|} \left[ \frac{e^{2}}{4\pi\varepsilon_{0}} \frac{4\pi}{N\Omega} \frac{\hbar\omega_{\mathrm{LO}}}{2} \left( \frac{1}{\epsilon_{\infty}} - \frac{1}{\epsilon_{0}} \right) \right]^{1/2}
	\end{equation}
	where $\mathbf{q}$ is the phonon wave vector, $\Omega$ is the unit cell volume, $N$ is the number of unit cells in the Born–von K\'{a}rm\'{a}n supercell, $\omega_{\mathrm{LO}}$ is the longitudinal optical phonon frequency, $e$ is the elementary charge, $\hbar$ is the reduced Planck constant, $\varepsilon_{0}$ is the vacuum permittivity, and $\epsilon_{\infty}$ and $\epsilon_{0}$ are the high-frequency and static dielectric constants, respectively.
	
	For the sake of convenience, the Fr\"{o}hlich interaction is commonly expressed through the dimensionless polar coupling constant $\alpha_{\mathrm{po}}$, which quantifies the overall strength of carrier coupling to LO phonons~\cite{ganose2021efficient}:
	\begin{equation}
		\alpha_{\mathrm{po}} = \frac{e^2}{4\pi\hbar\varepsilon_0} \left( \frac{m^{*}}{2\hbar\bar{\omega}} \right)^{1/2} \left( \frac{1}{\epsilon_{\infty}} - \frac{1}{\epsilon_{0}} \right)
	\end{equation}
	where $m^*$ is the carrier effective mass and $\bar{\omega}$ is the weighted longitudinal optical phonon frequency.
	
	To evaluate the POP scattering across layered materials, the crystal orientation of all the quantities entering $\alpha_{\mathrm{po}}$ must be considered, as electrical transport and dielectric response in layered structures differ substantially between in-plane and cross-plane directions. Since the cross-plane direction is the primary contributor to enhanced TE performance in layered systems—owing to its intrinsically low $\kappa_{\mathrm{L}}$—we compute $m^*$, $\epsilon_{\infty}$, and $\epsilon_{0}$ specifically along this direction.

	\vspace{0.3cm}
	\noindent \textbf{High Throughput Computational Screening.}
	As illustrated in Figure~\ref{fig1}, we selected 22 representative layered prototype structures from the Materials Project database~\cite{jain2013commentary} and identified 236 experimentally synthesized, non-magnetic semiconductors. In principle, the intrinsic TE transport properties of these materials can be evaluated by combining DFPT with electron Boltzmann transport theory. However, such calculations are computationally prohibitive within standard DFT frameworks, precluding their application in large-scale screening. Furthermore, many of these compounds are unlikely to exhibit competitive TE performance. To address this challenge, we developed a multistep high-throughput screening workflow that integrates DFT-based electronic structure calculations with physically motivated microscopic descriptors rooted in chemical bonding and dielectric response. We further incorporate explicit transport metrics---including carrier mobility $\mu$ and power factor PF---to progressively isolate compounds with superior intrinsic TE transport properties.
	
	To ensure computational feasibility while balancing the competing requirements of $\sigma$ and $S$ with respect to the effective mass in layered thermoelectric compounds, which typically exhibit low valley degeneracy along the cross-plane direction, we adopted a threshold of $m_{\mathrm{c}}^* \leq 0.8\,m_0$. At the same time, to avoid strong coupling between lattice distortions and electrons, i.e., polaron formation, which leads to substantial effective mass renormalization and reduced mobility, we employed $\alpha_{\mathrm{po}} = 0.8$ as a conservative upper bound to exclude materials prone to strong polaron formation.
	The first filter prioritizes favorable electrical transport by evaluating the cross-plane effective mass $m_{\mathrm{c}}^{}$ from self-consistent band structure calculations, retaining only materials with $m_{\mathrm{c}}^{} \le 0.8,m_0$.
	In parallel, we assessed the ionic contribution to the dielectric response, $\epsilon_{\mathrm{ion}} = \epsilon_{0} - \epsilon_{\infty}$, which directly governs the strength of POP scattering via the Fr\"{o}hlich coupling constant $\alpha_{\mathrm{po}}$. Accurate evaluation of the high-frequency dielectric constant $\epsilon_{\infty}$ is critical, as standard semilocal exchange--correlation functionals (e.g., PBEsol) tend to overestimate $\epsilon_{\infty}$ in narrow-gap semiconductors, thereby artificially reducing $\alpha_{\mathrm{po}}$. To mitigate this systematic error, we recalculated $\epsilon_{\infty}$ using the screened hybrid functional HSE06 under a finite electric-field perturbation, yielding more reliable dielectric screening and robust $\epsilon_{\infty}$ values. Using these corrected quantities, we computed the cross-plane $\alpha_{\mathrm{po}}$ for each candidate. 
	%Materials with $\alpha_{\mathrm{po}} \le 0.8$ were retained, as this regime signifies intrinsically weak Fr\"{o}hlich interactions conducive to high cross-plane $\mu$. 
	To bridge these microscopic descriptors with realistic transport performance, we explicitly evaluated the room-temperature cross-plane $\mu$ for all candidates satisfying the $m_{\mathrm{c}}^{*}$ and $\alpha_{\mathrm{po}}$ criteria, imposing a cutoff of $\mu \geq 300\,\mathrm{cm^{2}\,V^{-1}\,s^{-1}}$.
	This step removes compounds where carrier transport remains intrinsically limited despite favorable band dispersion and dielectric screening. Finally, we combined the recalculated $\mu$---accounting for ionized impurity (IMP) scattering---with the $S$ to determine the PF across different carrier concentrations at room temperature. Only materials achieving a maximum PF $\geq 20\,\mu\mathrm{W\,cm^{-1}\,K^{-2}}$ were selected. 
	This final threshold filters out systems where low $S$ or unfavorable band degeneracy limits the PF despite high $\mu$, ensuring that the surviving candidates simultaneously satisfy the requirements for efficient charge transport and strong TE performance.
	
	Figure~\ref{fig2} displays the distribution of all compounds in the parameter space defined by cross-plane $\alpha_{\mathrm{po}}$ and $m_{\mathrm{c}}^{*}$. The 42 compounds satisfying our screening criteria are highlighted in the inset. These two descriptors jointly quantify the interplay between band transport (favored by small $m_{\mathrm{c}}^{*}$) and polar phonon scattering (minimized by small $\alpha_{\mathrm{po}}$), both of which are essential for high $\mu$. The resulting materials landscape reveals two distinct categories. The first group, comprising Mg$_3$Sb$_2$- and SrIn$_2$P$_2$-type Zintl phases, occupies the low-$m_{\mathrm{c}}^{*}$ region, where steep band dispersion and anisotropic single-valley characteristics dominate electrical transport. In these materials, high $\mu$ arises primarily from favorable electronic structure rather than reduced POP scattering; their relatively large $\alpha_{\mathrm{po}}$ values reflect substantial Fr\"{o}hlich interactions that remain the principal mobility-limiting factor. In contrast, the second group consists mainly of layered transition-metal chalcogenides, such as PtS$_2$- and WSe$_2$-type systems, which reside near the lower bound of $\alpha_{\mathrm{po}}$. These systems exhibit intrinsically weak ionic polarization and minimal LO-phonon-induced macroscopic fields, resulting in POP scattering that approaches its theoretical minimum. Consequently, their mobility is governed not by ultralow $m_{\mathrm{c}}^{*}$, but by the suppression of the dominant long-range electron–phonon coupling. Together, these categories underscore complementary pathways toward high TE performance: the optimization of band dispersion to reduce $m_{\mathrm{c}}^{*}$, and the minimization of Fr\"{o}hlich interactions to suppress POP scattering. 
	
\begin{figure*}[tph!]
		\includegraphics[width=1.0\linewidth]{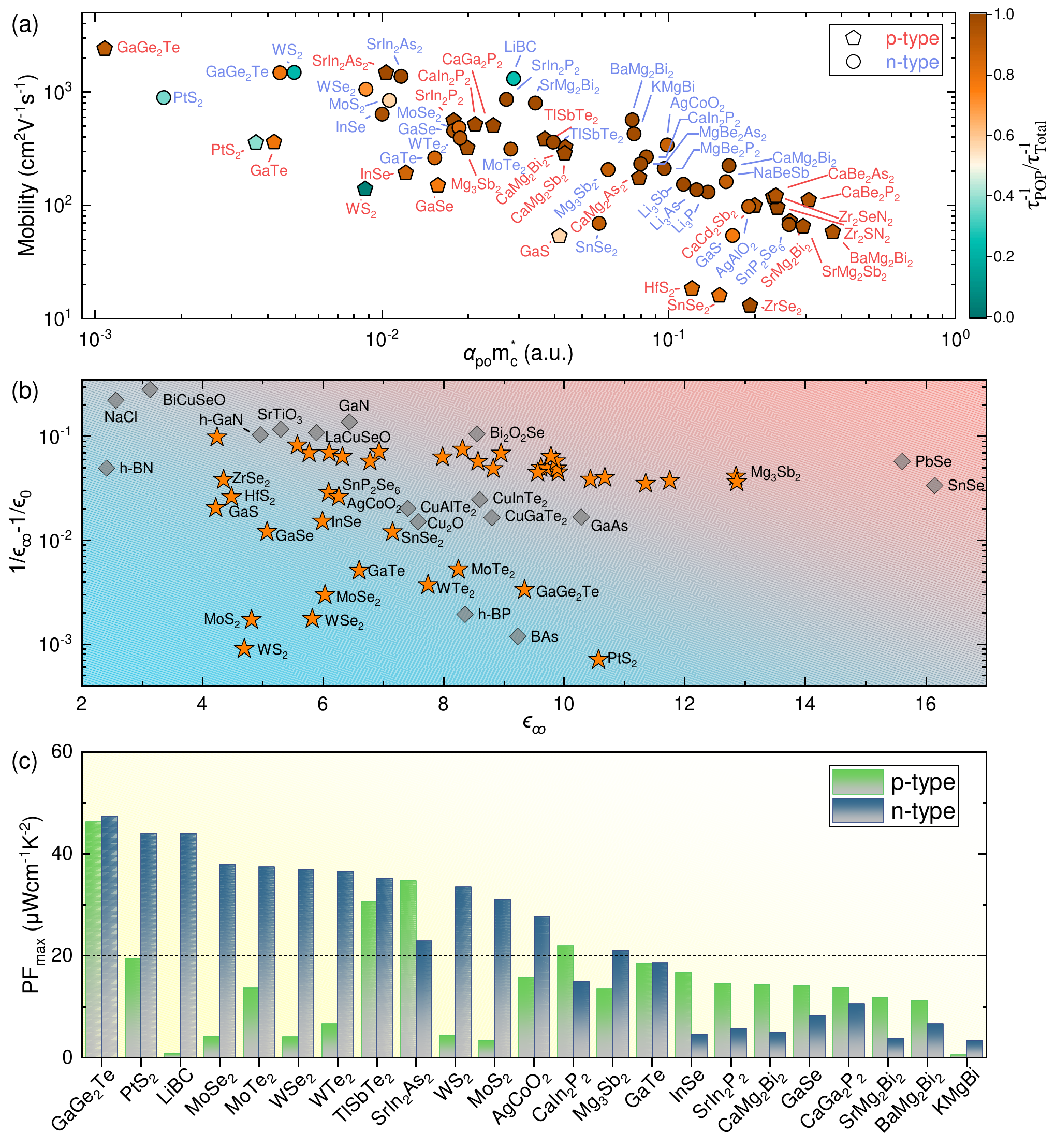}
		\caption{(a) The cross-plane carrier mobility at 300 K and a carrier concentration of 1 $\times$ 10$^{19}$ cm$^{-3}$, calculated by considering both ADP and POP scattering, as a function of the product of $\alpha_{\mathrm{po}}$ and $m_{\mathrm{c}}^{*}$ for materials that satisfy both screening thresholds. The symbol color represents the fraction of polar optical phonon scattering in the total scattering rate. (b) The correlation between the high-frequency dielectric constant $\epsilon_{\infty}$ and ${1}/{\epsilon_{\infty}} - {1}/{\epsilon_{0}}$. (c) The maximum power factor at room temperature among the high-mobility samples, considering three scattering mechanisms.}
		\label{mobility}
	\end{figure*}

	\vspace{0.3cm}
	\noindent \textbf{Origin of Low Polar Optical Phonon Scattering.} Our calculated $\mu$, accounting exclusively for phonon scattering mechanisms (ADP and POP), are presented in Figure~\ref{mobility}(a). At room temperature and a carrier concentration of $1\times10^{19}\,\mathrm{cm^{-3}}$, the majority of the shortlisted compounds exhibit relatively high $\mu$, consistent with their positioning in the low-$\alpha_{\mathrm{po}}$ and low-$m_{\mathrm{c}}^{*}$ region of the design map. A clear monotonic trend is observed where $\mu$ decreases with increasing $\alpha_{\mathrm{po}} m_{\mathrm{c}}^{*}$, confirming that transport limits in these systems are jointly governed by band transport (characterized by $m_{\mathrm{c}}^{*}$) and Fr\"{o}hlich coupling strength $\alpha_{\mathrm{po}}$. Notably, compounds with comparable $\alpha_{\mathrm{po}} m_{\mathrm{c}}^{*}$ values can exhibit substantially different $\mu$, reflecting the influence of secondary factors such as valley multiplicity, deformation potentials, elastic tensors, and the relative weighting of nonpolar scattering channels. Among the candidates, GaGe$_2$Te is distinguished by a remarkably high cross-plane hole mobility of $2280~\mathrm{cm^{2}\,V^{-1}\,s^{-1}}$ (ADP $+$ POP). Note that the calculated mobility values reported in this work represent only the intrinsic upper limits achievable under ideal conditions. In practice, approaching these values may require specialized synthesis techniques and careful carrier-concentration optimization. This value surpasses that of well-known high-mobility layered chalcogenides, such as PtS$_2$ and WSe$_2$, calculated at the same level of theory, despite the latter also possessing extremely low $\alpha_{\mathrm{po}}$. 
	However, when IMP scattering is included, the total mobility decreases to $645~\mathrm{cm^{2}\,V^{-1}\,s^{-1}}$. This substantial reduction from the intrinsic upper limit highlights a distinctive feature of GaGe$_2$Te: its intrinsically weak electron–phonon coupling renders IMP scattering, rather than POP scattering, a non-negligible mobility-limiting mechanism at high carrier concentrations.
	
	The color scale in Figure~\ref{mobility}(a) represents the fractional contribution of POP scattering to the total electron--phonon scattering rate. For most compounds, POP scattering dominates the relaxation process, frequently accounting for over 80\% of the total rate. This confirms that in polar or partially ionic layered materials, the long-range Fr\"{o}hlich interaction remains the primary mobility-limiting mechanism. Exceptions to this trend include LiBC, WS$_2$, and PtS$_2$, which display unusually weak POP contributions. In LiBC, this behavior stems from exceptionally high optical phonon frequencies—a consequence of light constituent atoms—yielding a large $\omega_{\mathrm{LO}}$ that suppresses the Fr\"{o}hlich coupling constant $\alpha_{\mathrm{po}}$. Conversely, the weak POP signature in PtS$_2$ and WS$_2$ arises from extremely flat cross-plane electronic bands. The resulting low carrier group velocities along this direction intrinsically limit the efficacy of long-range POP scattering, thereby allowing ADP scattering to dominate relaxation dynamics. Collectively, these results suggest that realizing high-performance TE requires the fundamental suppression of POP scattering, rather than reliance solely on band structure features such as $m_{\mathrm{c}}^{*}$ anisotropy or valley multiplicity. In this context, GaGe$_2$Te represents an optimal scenario where intrinsically weak Fr\"{o}hlich interaction synergizes with moderately dispersive bands to yield exceptionally high $\mu$. The high $\mu$ of GaGe$_2$Te arises from a combination of a highly dispersive band along the cross-plane direction, which gives rise to a small effective mass, and intrinsically weak POP scattering due to its small ionic dielectric constant $\epsilon_{\mathrm{ion}}$. This mechanism becomes particularly clear when GaGe$_2$Te is compared with other high-$\mu$ layered chalcogenides, such as WSe$_2$ and InSe. In WSe$_2$, the purely vdW interlayer bonding results in very flat cross-plane bands and a relatively heavy effective mass (0.44 $m_0$), despite weak POP scattering associated with its unusually small ionic dielectric constant ($\epsilon_{\mathrm{ion}}^{\parallel}=0.06$). By contrast, InSe possesses a three-dimensional interconnected network that leads to a light effective mass (0.08 $m_0$), but its large Born effective charges and a high ionic dielectric response ($\epsilon_{\mathrm{ion}}^{\parallel}=0.6$) lead to strong POP scattering that limits its mobility at room temperature. GaGe$_2$Te uniquely bridges these two extremes: the intercalated Ge bilayer induces strong $p_z$ orbital hybridization between layers, reducing the cross-plane effective mass to 0.05 $m_0$, while its robust covalent bonding suppresses the ionic dielectric response ($\epsilon_{\mathrm{ion}}^{\parallel}=0.3$), thereby rendering POP scattering negligible. As a result, GaGe$_2$Te effectively avoids the usual trade-off between effective mass and POP scattering, setting it apart from conventional layered semiconductors.
	
	To elucidate the dielectric origin of the suppressed Fr\"{o}hlich interaction, Figure~\ref{mobility}(b) presents the correlation between the high-frequency dielectric constant $\epsilon_{\infty}$ and the screening difference $\left( 1/\epsilon_{\infty} - 1/\epsilon_{0} \right)$, which directly governs the strength of long-range polar coupling. As $\epsilon_{\infty}$ increases, the quantity $\left( 1/\epsilon_{\infty} - 1/\epsilon_{0} \right)$ systematically decreases, indicating reduced lattice polarization and a weaker macroscopic electric field associated with LO phonons. Compounds with small $\left( 1/\epsilon_{\infty} - 1/\epsilon_{0} \right)$ predominantly cluster within the $\epsilon_{\infty}$ range of 4--11, coinciding with the regime of minimal POP scattering and maximal carrier mobility in Figure~\ref{mobility}(a). This trend demonstrates that the mitigation of Fr\"{o}hlich coupling is governed primarily by a reduced ionic dielectric contribution $\epsilon_{\mathrm{ion}}$, rather than by enhanced electronic screening alone. Representative examples include strongly covalent layered semiconductors such as h-BAs, h-BP, and transition-metal dichalcogenides. Notably, several high-performance thermoelectric materials (e.g., PbTe, SnSe, and Mg$_3$Sb$_2$) also fall within this low-$\left( 1/\epsilon_{\infty} - 1/\epsilon_{0} \right)$ regime, primarily owing to their large high-frequency dielectric constants $\epsilon_{\infty}$ rather than a suppressed ionic contribution.
	
	Having established $\epsilon_{\mathrm{ion}}$ as a key dielectric descriptor of POP scattering strength, we further examine its chemical origin. A clear correlation emerges between low cross-plane $\epsilon_{\mathrm{ion}}$ and strong intralayer covalency, quantified by the integrated crystal orbital bond index (ICOBI) (Figure~\textcolor{red}{S3}). High-mobility systems, including MoS$_2$-, PtS$_2$-, and GaGe$_2$Te-type structures, consistently exhibit small $\epsilon_{\mathrm{ion}}$ and large ICOBI values. In contrast, more ionic systems such as Zintl phases (e.g., Mg$_3$Sb$_2$ and SrIn$_2$P$_2$) show larger $\epsilon_{\mathrm{ion}}$ and weaker covalent bonding. Microscopically, strong covalent bonding implies substantial interatomic orbital overlap and electron delocalization. Upon ionic displacement, such delocalized electronic states undergo only limited charge redistribution, leading to reduced Born effective charges and weakened macroscopic polarization. Consequently, the Fr\"{o}hlich coupling constant $\alpha_{\mathrm{po}}$ is intrinsically suppressed, resulting in diminished POP scattering and enhanced carrier mobility.

	Importantly, this trend is independent of specific structural motifs. Across diverse compositions—from transition-metal dichalcogenides to tetrahedral layered semiconductors—compounds with significant interatomic orbital overlap systematically display reduced $\epsilon_{\mathrm{ion}}$ and weaker POP scattering. This observation highlights chemical bonding engineering, specifically the enhancement of covalent character along the transport direction, as an effective strategy for optimizing $\mu$ in layered TE materials.
	
	Figure~\ref{mobility}(c) summarizes the maximum power factor ($\mathrm{PF}_{\max}$) calculated within the carrier concentration range of $1.0 \times 10^{18}$ to $1.0 \times 10^{20}$~cm$^{-3}$ at room temperature. Fourteen compounds appear with $\mathrm{PF}_{\max}$ values exceeding $20~\mu\mathrm{W\,cm^{-1}\,K^{-2}}$, indicating a favorable tradeoff between high $\sigma$ and sufficiently large $S$. GaGe$_2$Te stands out by achieving exceptionally high $\mathrm{PF}_{\max}$ in both $p$-type and $n$-type regimes. This reflects its balanced and moderately dispersive band structure, which facilitates efficient transport for both carrier polarities. Several layered chalcogenides, including PtS$_2$, MoSe$_2$, WSe$_2$, and WTe$_2$, also display relatively large $\mathrm{PF}_{\max}$, particularly in the $n$-type regime, benefiting from the combination of weak POP scattering and moderate $m_{\mathrm{c}}^{*}$. Conversely, compounds with lower $\mathrm{PF}_{\max}$, despite verifying the $\mu$ screening criteria, often suffer from limited $S$ or insufficient band degeneracy. Thus, while high $\mu$ is a prerequisite, it is not solely sufficient for superior TE performance.
	
	In summary, integrating mobility-driven screening with explicit power-factor evaluation establishes a physics-based framework for identifying intrinsically high-performance thermoelectric materials. Among the 14 promising candidates identified through this workflow, GaGe$_2$Te emerges as the most compelling prototype for validating our design principle. Although several compounds exhibit competitive power factors, GaGe$_2$Te uniquely exemplifies the targeted microscopic mechanism: the incorporation of a covalent Ge bilayer intrinsically reduces the ionic dielectric response, thereby suppressing polar optical phonon scattering through weakened Fr\"{o}hlich coupling. Beyond its optimized electronic transport, GaGe$_2$Te also offers a structural advantage. Unlike high-mobility Zintl phases, which often exhibit relatively high lattice thermal conductivity, its van der Waals layered architecture suggests intrinsically suppressed phonon transport and the potential for ultralow $\kappa_\mathrm{L}$. Notably, both $n$-type and $p$-type transport display comparably high mobilities and power factors, indicating that the suppression of polar optical phonon scattering benefits electrons and holes alike. This rare synergy between reduced carrier scattering and favorable lattice dynamics positions GaGe$_2$Te as an ideal model system for elucidating the interplay between covalency engineering and thermoelectric performance. Given its unique combination of microscopic design principles and macroscopic transport advantages, we selected GaGe$_2$Te as the primary system for comprehensive first-principles investigation and benchmarked its transport behavior against the extensively studied Mg$_3$Sb$_2$ system. In addition, three other high $\mathrm{PF}$ candidates---PtS$_2$, TlSbTe$_2$, and SrIn$_2$As$_2$---were examined for comparison. These materials represent distinct electronic and bonding environments. PtS$_2$, a layered dichalcogenide, exhibits a near-zero $\alpha_{\mathrm{po}}$ and therefore serves as a representative system with intrinsically weak POP scattering. SrIn$_2$As$_2$ possesses the lightest $m_{\mathrm{c}}^*$ among the screened compounds, which is expected to strongly suppress ADP scattering and favor high $\mu$. In contrast, TlSbTe$_2$ displays a similar $\alpha_{\mathrm{po}}$ to SrIn$_2$As$_2$ but a substantially heavier $m_{\mathrm{c}}^*$, which is anticipated to limit its electrical transport performance. By contrasting these systems with GaGe$_2$Te, we aim to clarify how variations in bonding characteristics, dielectric response, and band dispersion collectively influence carrier scattering mechanisms and ultimately determine thermoelectric transport properties.
	
	%	In summary, integrating mobility-based screening with explicit PF evaluation provides a robust strategy for identifying high-performance TE materials. Within this framework, GaGe$_2$Te emerges as an optimal candidate that simultaneously maximizes $\mu$ and PF, illustrating the efficacy of suppressing POP scattering through reduced ionic dielectric response and enhanced covalency. Motivated by its exceptional predicted performance, the following sections focus on the TE properties of GaGe$_2$Te as a representative high-efficiency, covalency-optimized material.

	\begin{figure}[tph!]
		\includegraphics[width=1\linewidth]{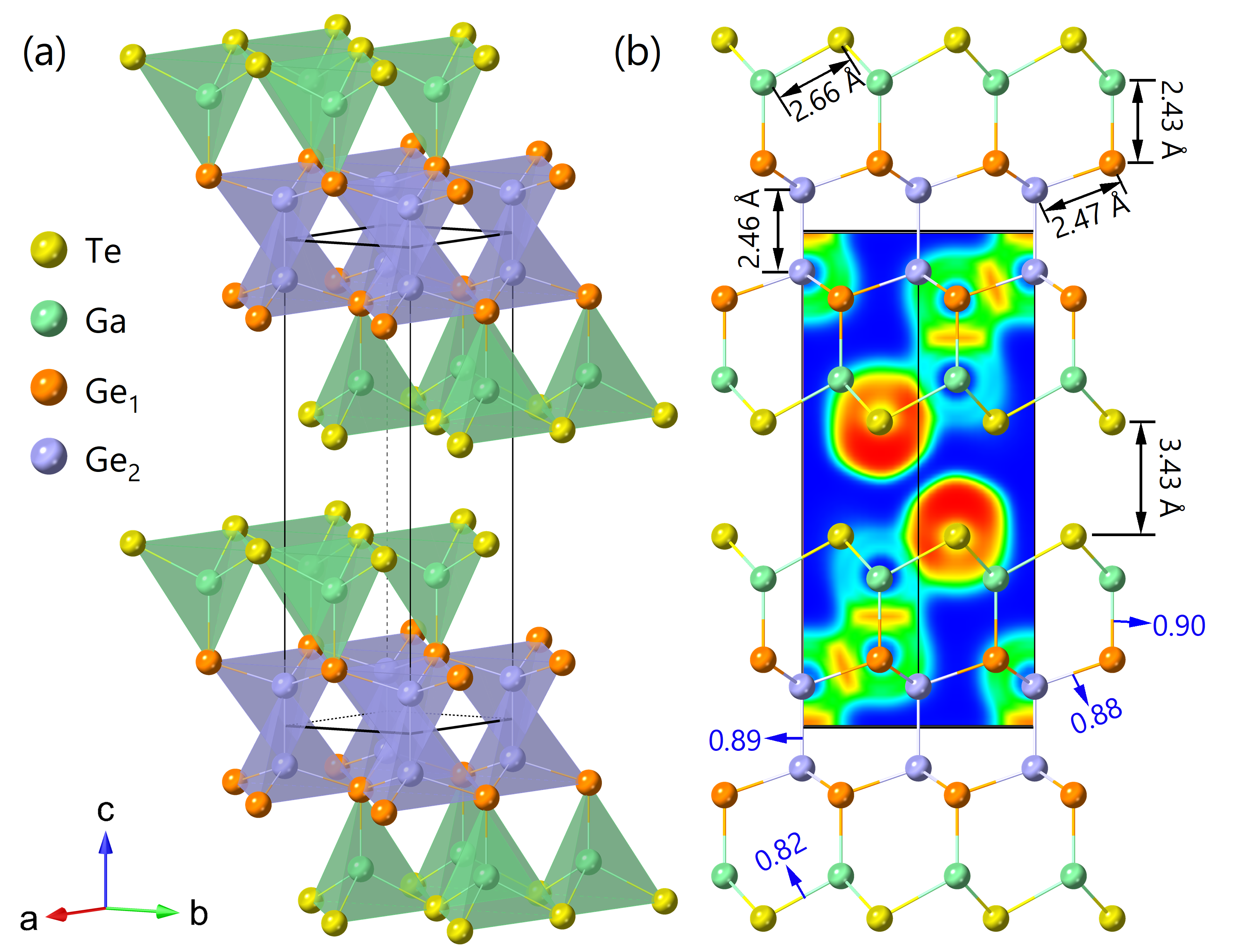}
		\caption{(a) The conventional unit cell and (b) the ELF of along [110] direction for GaGe$_2$Te. The black arrows mark the chemical bond lengths and the vertical distances between layers, and the blue arrows are the corresponding integrated crystal orbital bond index ICOBI.}
		\label{structure}
	\end{figure}

	\vspace{0.2cm}
	\noindent \textbf{Crystal structure and Chemical Bonds of GaGe$_2$Te.} GaGe$_2$Te crystallizes in the trigonal space group $P\bar{3}m1$ (No.~$164$). Our calculated lattice parameters ($a = b = 4.041$~\AA, $c = 14.935$~\AA) are in excellent agreement with experimental values~\cite{https://doi.org/10.1002/zaac.202300107}. As depicted in Figure~\ref{structure}, the crystal structure comprises covalently bonded slabs stacked along the $c$-axis, interacting via van der Waals (vdW) forces. Structurally, this phase can be rationalized as a derivative of $\beta$-GaGeTe formed by the insertion of an additional Ge atomic layer~\cite{gallego2022layered}, creating a motif where a Ge bilayer is sandwiched between Ga--Te blocks. The resulting interlayer spacing is $3.43$~\AA, which is comparable to that of the layered TE material Sb$_2$Si$_2$Te$_6$ ($3.15$~\AA)~\cite{LUO2020159} but notably larger than those of Bi$_2$Te$_3$ ($2.51$~\AA)~\cite{C6EE02017H} and TlCuSe ($2.54$~\AA)~\cite{https://doi.org/10.1002/adma.202104908}. This expanded interlayer separation indicates significantly weaker vdW coupling between adjacent slabs, promoting pronounced phononic anisotropy. Such decoupling naturally suppresses cross-plane heat conduction, a feature critical for minimizing $\kappa_{\mathrm{L}}$ and maximizing $\mathrm{ZT}$.
	
	Germanium atoms occupy two crystallographically distinct sites: Ge$_1$ (Wyckoff position $2d$) and Ge$_2$ ($2c$). The Ge$_2$ atoms function as tetrahedral centers, bonding to three Ge$_1$ atoms and one neighboring Ge$_2$ atom. The calculated Ge--Ge bond lengths ($2.468$~\AA and $2.459$~\AA) are remarkably close to those in elemental germanium ($2.450$~\AA), suggesting that the intercalated Ge bilayer preserves the bonding topology characteristic of a germanene-like network despite its confinement within the bulk framework. Additionally, Ga is tetrahedrally coordinated by one Ge$_1$ and three Te atoms. The Ga--Ge$_1$ and Ga--Te bond lengths ($2.434$~\AA and $2.660$~\AA, respectively) closely match the sums of their respective covalent single-bond radii~\cite{pyykko2009molecular}, reinforcing the predominantly covalent nature of the intralayer interaction network.
	
	This structural assignment is corroborated by the electron localization function (ELF), which reveals pronounced localization along Ga--Ge and Ge--Ge bonding paths, confirming strong directional covalency. Conversely, the interlayer regions exhibit near-zero ELF values, consistent with the weak vdW coupling inferred from geometric data. ICOBI analysis offers a quantitative perspective on these interactions. High ICOBI values are observed for Ge$_1$--Ge$_1$ (ICOBI $= 0.89$), Ge$_1$--Ge$_2$ (ICOBI $= 0.88$), and Ge$_1$--Ga (ICOBI $= 0.90$), signifying robust covalent bonding within the structural subunits. The Ga--Te interaction also displays substantial covalency (ICOBI $= 0.82$). These values approach those of archetypal covalent bonds (e.g., C--C in diamond, ICOBI $\approx 0.95$) and far exceed those of ionic systems such as NaCl (ICOBI $\approx 0.09$) calculated within the same framework. Critically, the strong Ge$_1$--Ge$_2$ coupling serves as evidence that the inserted Ge bilayer is not electronically inert but actively participates in the extended covalent network. Ultimately, this high degree of covalency minimizes the ionic contribution to lattice dynamics, providing the structural underpinning for the suppressed Fr\"{o}hlich coupling and the exceptional $\mu$ discussed in the preceding section.

	\begin{figure*}[tph!]
		\includegraphics[width=1.0\linewidth]{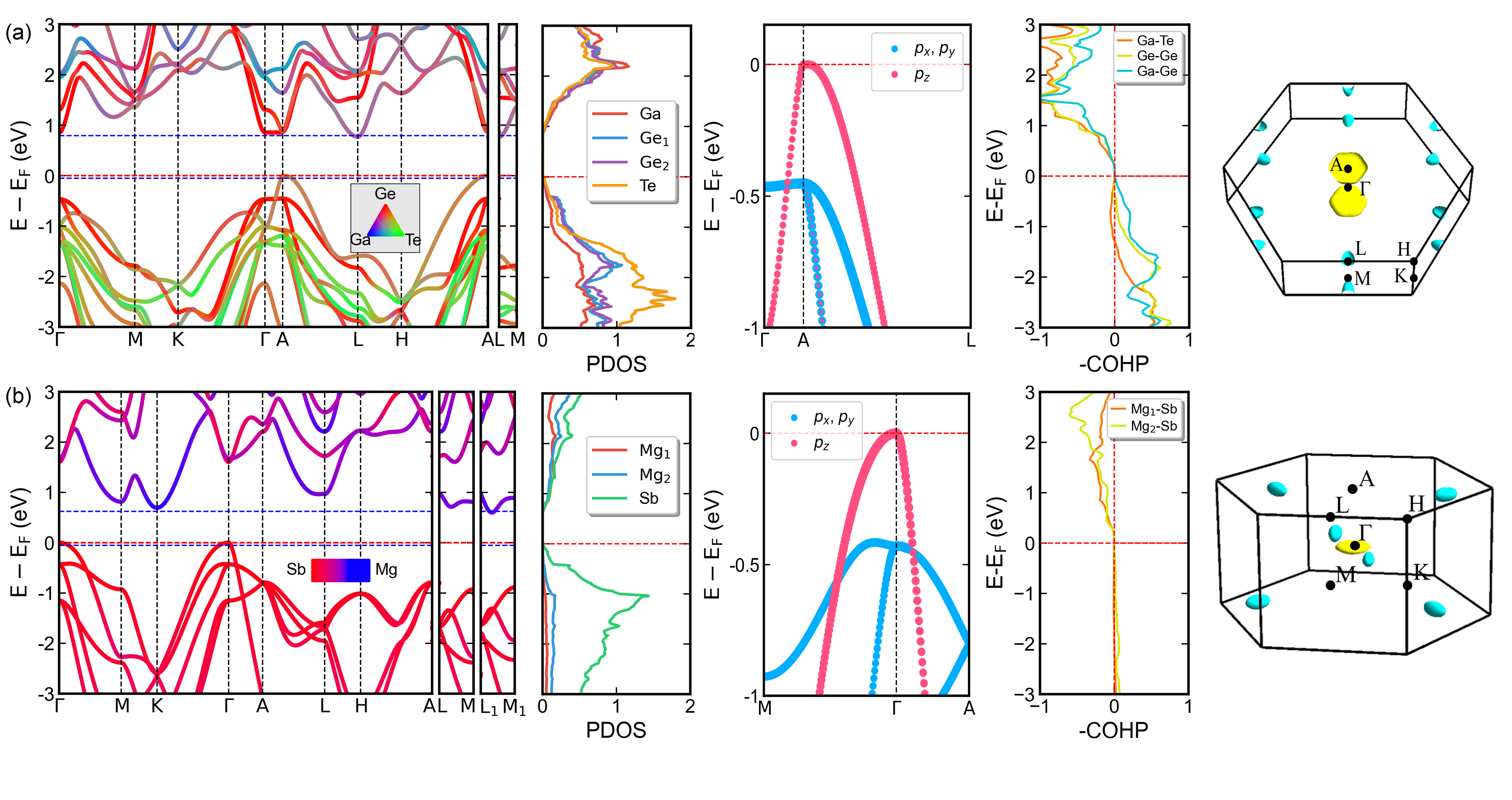}
		\caption{Element-resolved band structures, atom-projected density of states, enlarged views of the orbital-projected valence bands near the $\Gamma$ point, -COHP curves calculated using the PBEsol exchange–correlation functional, and Fermi surfaces of (a) GaGe$_2$Te and (b) Mg$_3$Sb$_2$. The Fermi surfaces corresponding energy levels are indicated by blue dashed lines in the band-structure panels. The yellow and cyan surfaces represent the Fermi surfaces near the VBM and CBM, respectively.}
		\label{band}
	\end{figure*}
	
	\vspace{0.2cm}
	\noindent \textbf{Electronic structures of GaGe$_2$Te.} An accurate description of the electronic band structure is a prerequisite for the reliable prediction of phonon-limited electrical transport properties. Figure~\ref{band} presents the electronic band structure of GaGe$_2$Te calculated using the HSE06 hybrid functional. Although HSE06 yields a substantially larger band gap ($0.77$~eV) compared to the PBEsol result ($0.13$~eV), the band dispersion remains largely unaffected. This indicates that the primary effect of the hybrid functional is a rigid upward shift of the conduction bands rather than a fundamental modification of the band curvature. GaGe$_2$Te is an indirect-gap semiconductor with the valence-band maximum (VBM) located at the A point. This contrasts with the well-studied TE material Mg$_3$Sb$_2$, where the VBM resides at the $\Gamma$ point. While both materials exhibit a single principal valence band pocket, the distinct momentum-space locations of these pockets stem from fundamental differences in interlayer bonding. In Mg$_3$Sb$_2$, the ionic layers are weakly coupled, localizing the highest valence states at the zone center.
	These states are dominated by Sb $5p$ orbitals and exhibit predominantly non-bonding character, with minimal hybridization with Mg orbitals. The absence of significant orbital overlap accounts for the negligible COHP values observed between $-3$~eV and the the Fermi level in Figure~\ref{band}(b).
	Conversely, in GaGe$_2$Te, the inserted Ge bilayer creates strong interlayer covalent Ge--Ge interactions, significantly enhancing band dispersion along the $\Gamma$--A (cross-plane) direction. This increased interlayer hopping pushes the VBM to the Brillouin-zone boundary (A point). The COHP analysis further confirms the presence of significant bonding interactions between Ge$_2$ atoms across adjacent layers, providing a chemical basis for the pronounced cross-plane dispersion.
	
	This structural dimensionality is directly mirrored in the band effective mass $m_\mathrm{b}^{*}$. In Mg$_3$Sb$_2$, the hole $m_\mathrm{b}^{*}$ values derived from band curvature along the in-plane ($\Gamma$--M/K) and cross-plane ($\Gamma$--A) directions are approximately $0.92~m_0$ and $0.07~m_0$, respectively. GaGe$_2$Te exhibits even lighter holes, with $m_\mathrm{b}^{*}$ of $0.74~m_0$ along A--H(L) and a remarkably low $0.05~m_0$ along A--$\Gamma$. Such a small $m_\mathrm{b}^{*}$ along the stacking direction is the primary contributor to the low $m_\mathrm{c}^{*}$ used in our screening, which dictates the high cross-plane mobility. This extremely low cross-plane $m_\mathrm{b}^{*}$ is a direct consequence of the covalency-driven interlayer coupling. Projected density of states (PDOS) analysis reveals that while the VBM of Mg$_3$Sb$_2$ originates mainly from isolated Sb~$5p$ states, the VBM of GaGe$_2$Te comprises hybridized Te~$5p$ and Ge~$4p$ orbitals, reflecting stronger directional bonding.
	
	The conduction band landscapes of the two systems diverge even more sharply. The conduction-band minimum (CBM) of Mg$_3$Sb$_2$ lies along the low-symmetry L$_1$–M$_1$ line, generating six symmetry-equivalent electron pockets~\cite{zhang2017high}. This high $N_{\mathrm{v}}$ is central to the high $S$ observed in $n$-type Mg$_3$Sb$_2$. In contrast, the CBM of GaGe$_2$Te is located at the high-symmetry L point, resulting in a valley degeneracy of $N_{\mathrm{v}} = 3$. Although this intrinsic degeneracy is lower than that of Mg$_3$Sb$_2$, the convergence of adjacent bands significantly compensates for this deficit. The energy separation between the L and A valleys is less than $20$~meV, and secondary conduction band edges at $\Gamma$ and A lie in close energetic proximity. Consequently, at relevant doping levels, these additional valleys become accessible, effectively enhancing the total degeneracy to a level comparable to Mg$_3$Sb$_2$. Chemical bonding analysis indicates that these conduction states arise primarily from hybridized Ga~$4p$ and Ge~$4p$ orbitals. Notably, such states dispersing along the $\Gamma$--A direction exhibit pronounced Ge~$p_z$ character, consistent with the directional nature of the interlayer orbital mixing.
	
	To further elucidate the microscopic origin of the exceptional electronic transport properties of the screened candidates, we analyzed the element-resolved band structures and PDOS of PtS$_2$, TlSbTe$_2$, and SrIn$_2$As$_2$, as shown in Figure~\textcolor{red}{S4}. PtS$_2$ exhibits an indirect band gap, with conduction-band characteristics that closely resemble those of the high-performance $n$-type material Mg$_3$Sb$_2$. Specifically, the CBM is located along the low-symmetry L$_1$--M$_1$ line, resulting in an unusually high valley degeneracy ($N_{\mathrm{v}} = 6$). Importantly, this primary valley is nearly degenerate with a secondary pocket along the H--A direction. Such band convergence effectively enhances the density-of-states effective mass, thereby improving $S$. Meanwhile, the predominantly covalent bonding character gives rise to a small ionic dielectric response, which suppresses POP scattering and helps maintain high carrier mobility. In contrast, TlSbTe$_2$ and SrIn$_2$As$_2$ represent a different paradigm for achieving high mobility, characterized by direct and relatively narrow band gaps at the $\Gamma$ point. The strong coupling between valence and conduction bands in narrow-gap semiconductors normally gives rise to highly dispersive band edges and extremely light inertial effective masses. Notably, unlike GaGe$_2$Te or PtS$_2$, these compounds possess much larger ionic dielectric constants ($\epsilon_{\mathrm{ion}} = 9.2$ for SrIn$_2$As$_2$ and $\epsilon_{\mathrm{ion}} = 62.3$ for TlSbTe$_2$), indicating highly polarizable lattices and strong Fr\"{o}hlich electron--phonon interactions. Consequently, their high mobilities do not arise from suppressed POP scattering; rather, they originate primarily from their ultralight carrier effective masses, which compensate for the strong scattering rates. From an orbital perspective, these dispersive band edges facilitate efficient charge transport. In TlSbTe$_2$, the VBM is dominated by Te-$p$ states, whereas the CBM arises from strong Sb--Te hybridization. Similarly, the VBM of SrIn$_2$As$_2$ is composed almost entirely of As states, whereas the CBM receives significant contributions from both In and As orbitals. Together, these contrasting mechanisms clearly reflect the two distinct boundaries of our mobility screening map (Figure~\ref{fig2}), ultimately highlighting the unique structural synergy of GaGe$_2$Te, which successfully integrates both a light effective mass and intrinsically weak POP scattering.

	\begin{figure*}[tph!]
		\includegraphics[width=1.0\linewidth]{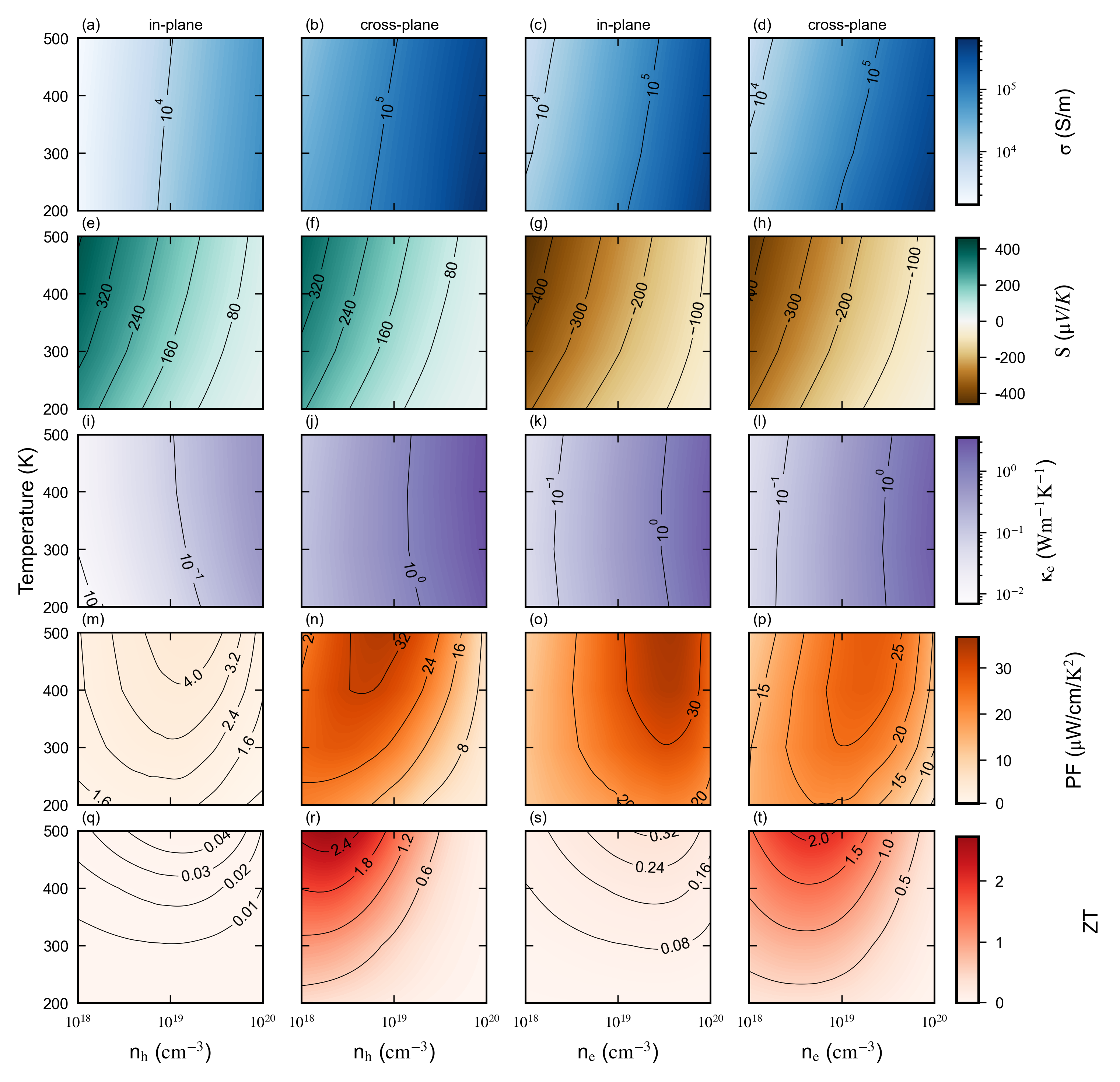}
		\caption{Calculated electronic transport properties of GaGe$_2$Te for $p$-type (left two columns) and $n$-type (right two columns) doping using the Perturbo package, including both electron--phonon and impurity (IMP) scattering: (a)--(d) electrical conductivity $\sigma$, (e)--(h) Seebeck coefficient $S$, (i)--(l) electrical thermal conductivity $\kappa_\mathrm{e}$, (m)--(p) power factor PF, and (q)--(t) figure of merit $\mathrm{ZT}$, for carrier concentrations from $10^{18}$--$10^{20}$ cm$^{-3}$ and temperatures from 200 to 500 K.}
		\label{PF}
	\end{figure*}

	\vspace{0.2cm}
	\noindent \textbf{Electrical transport properties and scattering rates of GaGe$_2$Te.} The electrical transport coefficients ($\sigma$, $S$, $\kappa_\mathrm{e}$, and PF) of GaGe$_2$Te were calculated as functions of temperature and carrier concentration for both $n$-type and $p$-type doping. Consistent with its layered crystal structure, GaGe$_2$Te exhibits pronounced transport anisotropy, particularly for $p$-type carriers. As established in the preceding electronic structure analysis, the hole $m_{\mathrm{c}}^{*}$ is highly anisotropic ($m_{\perp}^*/m_{\parallel}^* \approx 12.5$). This extremely light cross-plane $m_{\mathrm{c}}^{*}$ drives a significantly higher cross-plane $\sigma$ compared to the in-plane direction. Conversely, the electron $m_{\mathrm{c}}^{*}$ under $n$-type doping is nearly isotropic ($m_{\perp}^*/m_{\parallel}^* \approx 1$), leading to commensurate values for in-plane and cross-plane $\sigma$. Since $\kappa_\mathrm{e}$ is coupled to $\sigma$ via the Wiedemann–Franz law, $\kappa_\mathrm{e}=L \sigma \mathrm{T}$~\cite{ashcroft1979solid}, the anisotropy profile of $\kappa_\mathrm{e}$ mirrors that of $\sigma$. At high carrier concentrations, the cross-plane $\kappa_\mathrm{e}$ becomes comparable in magnitude to the $\kappa_\mathrm{L}$, a factor that reduces the ratio of electrical to thermal conductivity and imposes an intrinsic limit on the maximal figure of merit $\mathrm{ZT}$ at elevated doping levels.
	
	As expected, $S$ follows an inverse dependence relative to $\sigma$. Under $n$-type doping, $S$ reaches values of $456~\mu\mathrm{V\,K^{-1}}$ (in-plane) and $431~\mu\mathrm{V\,K^{-1}}$ (cross-plane) at a carrier concentration of $1\times10^{18}~\mathrm{cm^{-3}}$ and 500~K, surpassing the corresponding values for $p$-type doping ($406~\mu\mathrm{V\,K^{-1}}$ and $378~\mu\mathrm{V\,K^{-1}}$). This enhancement stems from the multi-valley character of the conduction band edge and the heavy-band features along the $\Gamma$–A direction, which increase the $m_{\mathrm{d}}^*$ and thereby boost $S$. The integration of these transport parameters yields PF. For $p$-type GaGe$_2$Te, the PF culminates in a maximum of $71.8~\mu\mathrm{W\,cm^{-1}\,K^{-2}}$ along the cross-plane direction at 200~K and hole concentration of $3\times10^{18}~\mathrm{cm^{-3}}$, far exceeding the in-plane value of $9.9~\mu\mathrm{W\,cm^{-1}\,K^{-2}}$. In the $n$-type regime, the interplay of slightly higher in-plane $\sigma$ and comparable $S$ results in a maximum PF of $55.2~\mu\mathrm{W\,cm^{-1}\,K^{-2}}$ at 200~K and electron concentration of $4\times10^{19}~\mathrm{cm^{-3}}$. To benchmark these results, the transport properties of Mg$_3$Sb$_2$ are presented in Figure~\textcolor{red}{S5}. While the high valley degeneracy of the CBM in Mg$_3$Sb$_2$ yields a superior $S$, the concomitant heavy $m_{\mathrm{c}}^{*}$ suppresses $\mu$, resulting in lower overall PF values for $n$-type Mg$_3$Sb$_2$ compared to the best performance of GaGe$_2$Te. Furthermore, $p$-type Mg$_3$Sb$_2$ suffers from a low $S$, yielding a PF approximately one-third of its $n$-type counterpart and significantly lower than that of $p$-type GaGe$_2$Te.
	Similarly, in PtS$_2$, the sixfold-degenerate valley at the CBM gives rise to a large $m_{\mathrm{d}}^*$. As a result, $n$-type PtS$_2$ exhibits a large and relatively isotropic $S$ (Figure~\textcolor{red}{S6}). Because the large $m_{\mathrm{d}}^*$ prevents $S$ from degrading rapidly with increasing doping, the $\mathrm{PF}$ increases monotonically with carrier concentration, requiring heavy doping ($>10^{20}\,\mathrm{cm^{-3}}$) to shift the Fermi level deeper into the conduction band and maximize thermoelectric performance. In contrast, the $p$-type $\mathrm{PF}$ is substantially lower, primarily because the valence band, dominated by S-$p$ orbitals, has a heavier effective mass that limits hole mobility. In TlSbTe$_2$ (Figure~\textcolor{red}{S7}), the highly dispersive states at the $\Gamma$ point endow the material with excellent carrier mobility and a high $\mathrm{PF}$. However, similar to PtS$_2$, maximizing the $\mathrm{PF}$ in TlSbTe$_2$ requires relatively high carrier concentrations. At a doping level of approximately $5\times10^{20}\,\mathrm{cm^{-3}}$, TlSbTe$_2$ reaches maximum $\mathrm{PF}$ values of about $60.6\,\mu\mathrm{W\,cm^{-1}\,K^{-2}}$ for $p$-type transport and $114.3\,\mu\mathrm{W\,cm^{-1}\,K^{-2}}$ for $n$-type transport along the cross-plane direction at 500 K. Similar to TlSbTe$_2$, SrIn$_2$As$_2$ exhibits a relatively small $m_{\mathrm{d}}^*$ because of the absence of the high valley degeneracy found in PtS$_2$, which intrinsically leads to a lower $S$ (Figure~\textcolor{red}{S8}). Physically, the small $m_{\mathrm{d}}^*$ causes the Fermi level to shift rapidly into the band upon doping, resulting in a pronounced reduction in $S$. Consequently, the optimal trade-off between $S$ and $\sigma$ occurs at carrier concentrations much lower than those required for PtS$_2$. At the same time, the ultralight effective mass still ensures sufficiently high $\sigma$. As a result, SrIn$_2$As$_2$ reaches its optimal power factor at a carrier concentration of approximately $10^{19}\,\mathrm{cm^{-3}}$. At room temperature, the maximum cross-plane power factor reaches about $35.7\,\mu\mathrm{W\,cm^{-1}\,K^{-2}}$ for $p$-type transport and $24.5\,\mu\mathrm{W\,cm^{-1}\,K^{-2}}$ for $n$-type transport.
	
	However, GaGe$_2$Te uniquely integrates several favorable characteristics of these systems. It combines the small effective mass observed in SrIn$_2$As$_2$ with the weak POP scattering associated with the small ionic dielectric response of PtS$_2$, while simultaneously supporting efficient $p$-type and $n$-type transport channels similar to those in TlSbTe$_2$. As a result, GaGe$_2$Te exhibits a high $\mathrm{PF}$ and can achieve excellent $\mathrm{ZT}$ values at lower and more experimentally accessible carrier concentrations.

	\begin{figure*}[tph!]
		\includegraphics[width=1.0\linewidth]{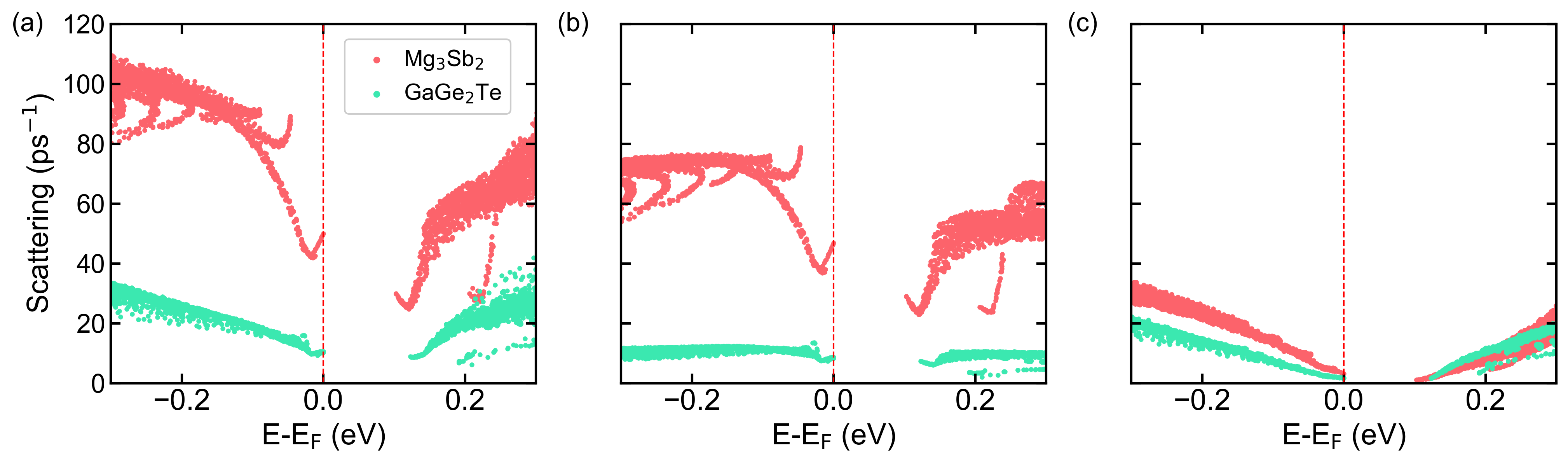}
		\caption{(a) Total scattering rate (b) long-range and (c) remainder contributions for $n$-type and $p$-type doping. }
		\label{scattering}
	\end{figure*}

	Figure~\ref{scattering} elucidates the carrier relaxation dynamics by comparing the electron--phonon scattering rates ($\tau^{-1}$) of GaGe$_2$Te and Mg$_3$Sb$_2$ at 300~K.
	The total scattering rate in Mg$_3$Sb$_2$ exhibits a steep divergence near the band edge, reaching values nearly four times higher than those of GaGe$_2$Te. Decomposition of the total scattering rate clearly shows that this transport bottleneck originates from long-range polar interactions rather than short-range deformation-potential scattering. In Mg$_3$Sb$_2$, the scattering rate follows the characteristic $1/k$ divergence associated with the Fr\"{o}hlich interaction, reflecting the dominance of long-range polar optical phonon scattering. This singular behavior strongly enhances small-momentum carrier scattering near the band edge, thereby severely limiting carrier mobility. In contrast, GaGe$_2$Te exhibits a flattened and strongly suppressed scattering profile in the low-$k$ region, indicating that the macroscopic electric field associated with long-wavelength LO phonons is substantially weakened. This behavior provides direct evidence of reduced Fr\"{o}hlich coupling arising from its diminished ionic dielectric response. Ultimately, this pronounced contrast can be traced to the fundamentally different dielectric origins of the two materials.
	Mg$_3$Sb$_2$ is a highly ionic Zintl compound with large Born effective charges, which generate strong macroscopic polarization fields upon LO phonon excitation. As shown in Figure~\textcolor{red}{S9}, these large dynamic charges drive pronounced LO--TO splittings, consistent with strong long-range Fr\"{o}hlich coupling. Conversely, GaGe$2$Te features predominantly covalent bonding within its layered network. Its Born effective charges are small, and its $\epsilon{\mathrm{ion}}$ is nearly an order of magnitude lower than that of Mg$_3$Sb$_2$. Consequently, LO--TO splitting is negligible and POP coupling to charge carriers is intrinsically weak, rendering long-range scattering naturally suppressed without extrinsic engineering. Crucially, Figure~\ref{scattering}(c) reveals that the residual short-range electron--phonon scattering rates remain comparable in the two systems. Therefore, the suppression of polarity in GaGe$_2$Te does not lead to enhanced short-range electron--phonon scattering. The material thus achieves a rare combination of weak long-range scattering and low short-range scattering, effectively avoiding the typical trade-off between polarity suppression and lattice softness.
	The suppression of Fr\"{o}hlich interaction translates directly into dramatically enhanced $\mu$. Despite possessing a comparable hole $m_{\mathrm{c}}^{*}$ to Mg$_3$Sb$_2$, GaGe$_2$Te exhibits a room-temperature hole mobility more than seven times higher ($2280~\mathrm{cm^2\,V^{-1}\,s^{-1}}$ vs. $306~\mathrm{cm^2\,V^{-1}\,s^{-1}}$). The exceptional transport in GaGe$_2$Te is therefore not primarily governed by band curvature, but fundamentally by its covalent bonding topology that minimizes ionic polarizability.
	
	In summary, GaGe$_2$Te synergizes light cross-plane hole $m_{\mathrm{c}}^{*}$ with intrinsically suppressed POP scattering to yield superior $\sigma$ and PF, particularly under $p$-type doping. For $n$-type doping, while the PF differs less from that of Mg$_3$Sb$_2$, the significantly lower cross-plane $\kappa_{\mathrm{L}}$ of the vdW layered structure positions GaGe$_2$Te for superior overall TE performance. These findings highlight the minimization of ionic dielectric response---via enhanced covalency---as a critical materials design principle for high-performance thermoelectrics.

	\begin{figure*}[tph!]
		\includegraphics[width=1.0\linewidth]{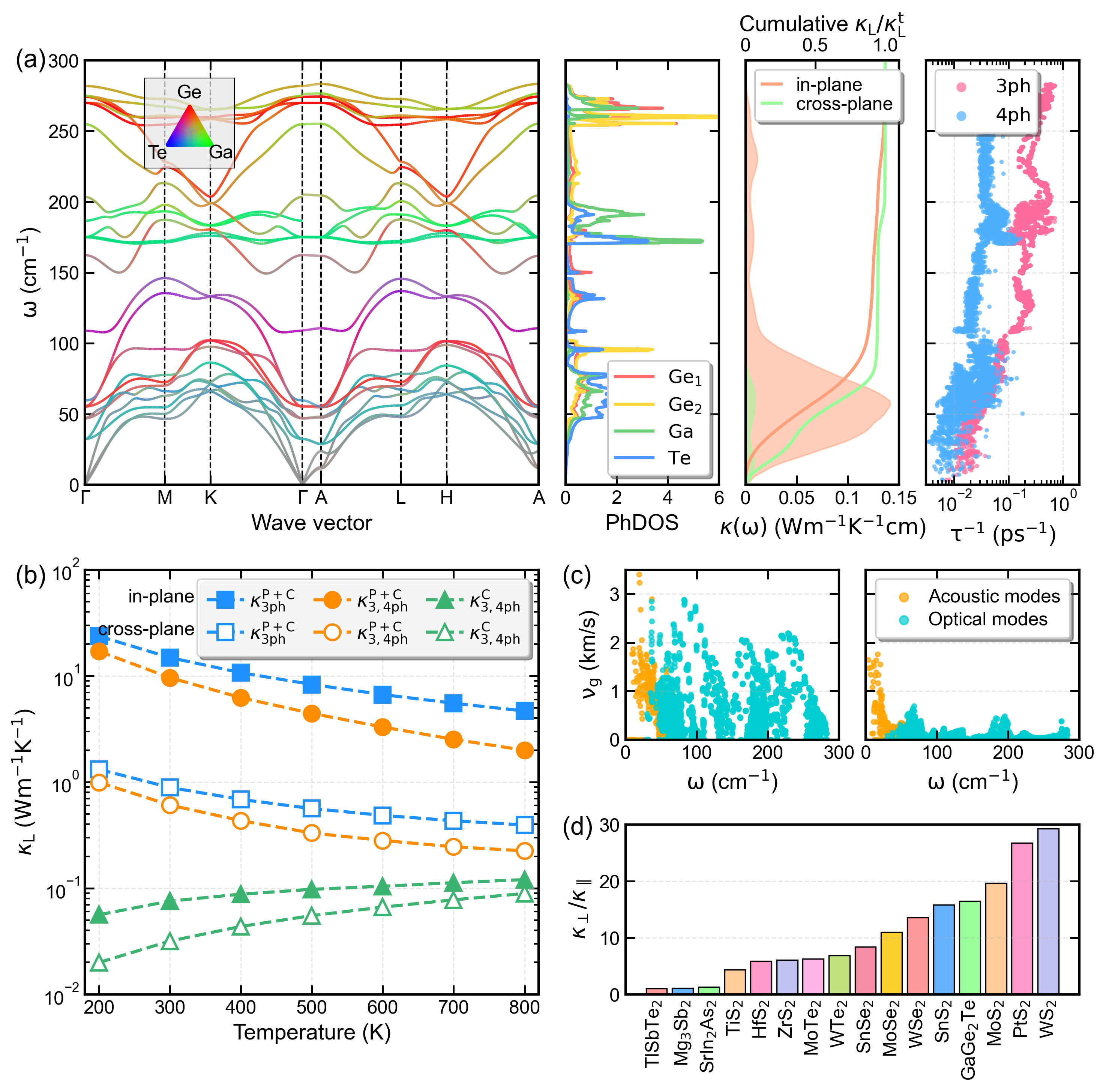}
		\caption{(a) The atom resolution phonon dispersion including non-analytical corrections, phonon density of states PhDOS, lattice conductivity spectrum $\kappa(\omega)$, the ratio cumulative $\kappa_{\mathrm L}$ to total lattice thermal conductivity $\kappa_{\mathrm L}^{\mathrm t}$ and the 3ph and 4ph scattering rates ($\tau^{-1}$) at 300 K. (b) The calculated $\kappa_{\mathrm L}$ as a function of temperature in in-plane and cross-plane directions. (c) The phonon group velocity at room temperature in in-plane (left) and cross-plane (right). (d) The $\kappa_{\perp}/\kappa_{\parallel}$ in different compounds\cite{li2021demonstration,lindroth2016thermal,10.1021/acs.chemmater.0c04184,snse2,glebko2019lattice}.}
		\label{fatphonon}
	\end{figure*}

	\vspace{0.2cm}
	\noindent \textbf{Lattice dynamics and phonon transport.}
	The phonon properties of GaGe$_2$Te were investigated to elucidate the origin of its ultralow cross-plane $\kappa_{\mathrm{L}}$. As depicted in Figure~\textcolor{red}{S1}, the phonon dispersion relations demonstrate no imaginary frequencies between 0--800 K, confirming dynamical stability at 0~K. 
	In addition to the dynamical stability confirmed by phonon dispersion, we comprehensively evaluated the thermodynamic, mechanical, and thermal stability of GaGe$_2$Te. Our calculated decomposing energy is -2.8 meV/atom, indicating its intrinsic thermodynamic stability at 0 K. The elastic constants satisfy the Born--Huang criteria for the trigonal crystal system (Table~\textcolor{red}{S2}), confirming the mechanical stability. Furthermore, high-temperature MD simulations performed at 800 K (Figure~\textcolor{red}{S2}) demonstrate that lattice framework remains intact, with no evidence of phase transition or decomposition. These results collectively attest to the structural integrity of GaGe$_2$Te within the temperature range relevant for thermoelectric applications.
	The atom-projected phonon dispersion and phonon density of states (PhDOS) of GaGe$_2$Te at 300 K are presented in Figure~\ref{fatphonon}(a). Given that the primitive unit cell of GaGe$_2$Te comprises 8 atoms (2 formula units), there are 24 phonon branches at the Brillouin zone center ($\Gamma$ point), which decompose as $\mathrm{4A_{1g}} \oplus \mathrm{4E_{g}} \oplus \mathrm{4A_{2u}} \oplus \mathrm{4E_{u}}$. Heavy Te atoms dominate the low-frequency phonon modes, whereas the lighter Ga and Ge atoms, possessing similar masses, govern the mid- and high-frequency branches. 
	
	The PhDOS indicates that Ge$_2$ atoms contribute more significantly than Ge$_1$ atoms. Notably, the flat band at $100~\mathrm{cm^{-1}}$ originates almost exclusively from Ge$_2$; located at both the center and vertices of the tetrahedron, Ge$_2$ atoms undergo cooperative vibrations analogous to rattling between the two layers of Ge$_1$ atoms. Near the Brillouin zone center ($\Gamma$ point), the degeneracy of the LO and TO phonons at $180~\mathrm{cm^{-1}}$ is lifted, indicating LO--TO splitting. The low $\epsilon_{\mathrm{ion}}$ of GaGe$_2$Te implies weak ionic polarization, resulting in ineffective screening of the long-range Coulomb field associated with LO phonons. Consequently, consistent with the Lyddane--Sachs--Teller (LST) relation, the material exhibits small LO--TO splitting. Since distinct optical phonon modes contribute unequally to the macroscopic electric polarization, the magnitude of the LO--TO splitting is mode-dependent. The largest splitting corresponds to the in-plane vibrational modes of Ga atoms, while smaller splittings at higher frequencies arise from the cross-plane vibrations of Ga and/or Ge atoms. The frequencies of the lowest-energy transverse optical (TO) phonon mode in GaGe$_2$Te are 33 cm$^{-1}$ and 29 cm$^{-1}$ at the $\Gamma$ and A points, respectively. The minimum phonon band gap between the acoustic and optical branches is only 5 cm$^{-1}$, which enables strong scattering between acoustic and optical phonons and thus provides abundant scattering channels for acoustic phonons, contributing to the ultralow $\kappa_{\mathrm{L}}$. It is worth noting that, unlike in Group IV--VI chalcogenides (e.g., SnSe, PbTe, and GeTe),~\cite{delaire2011giant,li2015orbitally} where low $\kappa_{\mathrm{L}}$ is associated with the lowest-energy TO modes and may lead to ferroelectric instabilities, local off-centering, and symmetry breaking, the lowest-energy TO phonon in GaGe$_2$Te remains dynamically stable even at 0 K.
	
	Weak chemical bonding, often accompanied by strong lattice anharmonicity, is a well-established precursor to low $\kappa_{\mathrm{L}}$~\cite{wang2025strong,xie2020first,wang2026atomic}. As a layered vdW compound, GaGe$_2$Te displays pronounced bonding heterogeneity, consisting of strong intralayer covalent bonds and weak interlayer vdW interactions. This bonding hierarchy gives rise to pronounced phonon dispersion anisotropy. In particular, the longitudinal acoustic (LA) phonon branch along the $\Gamma$--A (cross-plane) direction lies significantly lower than those along the $\Gamma$--K and $\Gamma$--M (in-plane) directions. Such reduced acoustic phonon frequencies indicate diminished phonon group velocities ($v_g$), which intrinsically suppress heat transport along the cross-plane direction.
	To further clarify the origin of the strong phonon scattering beyond the geometric effect of the layered structure, we calculated the mode-resolved Gr\"{u}neisen parameters ($\gamma$) of GaGe$_2$Te (Figure~\textcolor{red}{S10}). The acoustic branches, particularly along the $\Gamma$--A (cross-plane) direction, exhibit unusually large $\gamma$ values, indicating pronounced intrinsic anharmonicity. This enhanced anharmonicity arises from the strong bond heterogeneity between the rigid intralayer covalent bonding and the compliant interlayer vdW interactions. Consequently, both three-phonon and four-phonon scattering processes are significantly strengthened, leading to shortened phonon lifetimes and further suppression of the lattice thermal conductivity despite the robust in-plane covalent network.
	The $\kappa_{\mathrm{L}}$ of GaGe$_2$Te was calculated by incorporating phonon renormalization, three-phonon (3ph) and four-phonon (4ph) scattering processes, and coherence effects across a range of temperatures. As anticipated from its vdW layered structure, $\kappa_{\mathrm{L}}$ exhibits strong anisotropy: the in-plane $\kappa_{\mathrm{L}}$ reaches $9.6~\mathrm{W\,m^{-1}\,K^{-1}}$ at 300~K, approximately 16 times the cross-plane value of $0.57~\mathrm{W\,m^{-1}\,K^{-1}}$. Large thermal anisotropy ratios ($\kappa_{\perp}/\kappa_{\parallel}$) are characteristic of layered materials. For instance, at room temperature, ratios of 15.8, 13.5, 19.6, and 10.9 are observed in SnS$_2$~\cite{zhan2020phonon}, WSe$_2$, MoS$_2$, and MoSe$_2$~\cite{lindroth2016thermal}, respectively. This contrasts sharply with the nearly isotropic $\kappa_{\mathrm{L}}$ reported for Mg$_3$Sb$_2$-based compounds~\cite{li2021demonstration}. The substantial anisotropy ratio in GaGe$_2$Te is primarily driven by the enhanced in-plane $\kappa_{\mathrm{L}}$, which results from higher $v_g$ along the in-plane direction, as shown in Figure~\ref{fatphonon}(c). Ultimately, this thermal anisotropy is intrinsic to the chemical bonding environment, specifically the contrast between the strong covalent intralayer framework and the weak vdW interactions between adjacent layers. This structural motif effectively suppresses cross-plane phonon transport while facilitating efficient in-plane conduction.
	%	
	%	Figure~\ref{fatphonon} further illustrates the spectral contribution of phonons to the total $\kappa_{\mathrm{L}}$. Acoustic phonons and low-frequency optical phonons (below $100~\mathrm{cm^{-1}}$) account for over 90\% of $\kappa_{\mathrm{L}}$. The significant suppression of $\kappa_{\mathrm{L}}$ due to 4ph scattering also stems from this frequency range, which is primarily associated with the vibrations of Ga and Te atoms. 
	%	Furthermore, the analysis reveals that particle-like (diffusive) phonon transport is dominant, accounting for approximately 95\% of the total thermal conductivity even in the cross-plane direction. This finding suggests that wave-like (coherent or tunneling) transport channels play a negligible role in this compound.
	Figure~\ref{fatphonon} further illustrates the spectral contributions of phonons to the total $\kappa_{\mathrm{L}}$. Acoustic phonons and low-frequency optical phonons (below $100~\mathrm{cm^{-1}}$) account for more than 90\% of $\kappa_{\mathrm{L}}$. The pronounced suppression of $\kappa_{\mathrm{L}}$ arising from four-phonon (4ph) scattering also originates primarily from this frequency range, which is largely associated with the vibrations of Ga and Te atoms. Moreover, the analysis reveals that particle-like (diffusive) phonon transport dominates, contributing approximately 95\% of the total thermal conductivity, even along the cross-plane direction. This indicates that wave-like (coherent or tunneling) transport channels play a negligible role in this compound. Therefore, heat conduction is governed predominantly by intrinsic scattering processes rather than by coherent wave-like tunneling across the layers. The cross-plane thermal resistance thus arises from the combined effects of reduced phonon group velocities and shortened lifetimes, both of which are intrinsic to the bonding hierarchy of the material.
	
	Taken together, Figures~\ref{scattering} and ~\ref{fatphonon} establish a natural transport decoupling mechanism in GaGe$_2$Te: enhanced covalency suppresses long-range polar optical phonon scattering and improves carrier mobility, while the same layered bonding topology induces strong phonon softening and anharmonicity that minimize lattice thermal conductivity along the cross-plane direction. This rare synergy between electronic and phononic optimization underpins the superior thermoelectric performance predicted for this compound.

	\vspace{0.2cm}
	\noindent \textbf{Figure of merit $\mathrm{ZT}$.} 
	Before evaluating the thermoelectric performance of GaGe$_2$Te, we first benchmarked our computational approach using the well-studied thermoelectric material Mg$_3$Sb$_2$, for which extensive experimental transport data are available. As shown in Figure~\textcolor{red}{S11}, the $S$, $\mathrm{PF}$, and $\mathrm{ZT}$ values calculated from first-principles electron--phonon coupling combined with Boltzmann transport theory and IMP scattering theory are compared with representative experimental measurements for single-crystal Mg$_3$Sb$_2$.~\cite{li2021demonstration} Our calculations reproduce both the order of magnitude and the carrier-concentration dependence of the transport coefficients. In particular, the predicted $S$ and $\mathrm{PF}$ fall within the experimentally observed range at comparable carrier concentrations, yielding $\mathrm{ZT}$ values consistent with reported measurements. We note that experimental samples may involve extrinsic effects such as residual defects, dopants, or microstructural features, whereas the present calculations describe ideal bulk crystals and include only intrinsic electron--phonon scattering together with IMP scattering. Consequently, exact quantitative agreement is not expected. Nevertheless, the reasonable agreement in both magnitude and overall trends indicates that the present computational framework captures the dominant scattering mechanisms governing thermoelectric transport. On this basis, the same first-principles methodology was subsequently applied to GaGe$_2$Te without introducing any empirical fitting parameters.
%	Before evaluating the $\mathrm{ZT}$ performance of GaGe$_2$Te, we benchmarked our computational workflow against the well-studied Mg$_3$Sb$_2$ system, a material whose transport behavior is known to be accurately described by first-principles electron-phonon and Boltzmann transport calculations. The calculated $\mathrm{ZT}$ values for $n$-type and $p$-type Mg$_3$Sb$_2$ (Figure~\textcolor{red}{S5}) demonstrate excellent agreement with experimentally reported data. For $n$-type Mg$_3$Sb$_2$, our calculations yield nearly isotropic $\mathrm{ZT}$ values of 1.4 at 800~K at an optimal carrier concentration of $3\times 10^{19}\,\mathrm{cm}^{-3}$, consistent with the measured peak value of 1.85 at 723~K~\cite{chen2018extraordinary}. Similarly, the computed $p$-type $\mathrm{ZT}$ values--0.30 (in-plane) and 0.83 (cross-plane)--result in a spatially averaged $\mathrm{ZT}$ of 0.53 at 700~K. This is in close agreement with the experimentally observed $\mathrm{ZT} \approx 0.5$ for Ag-doped polycrystalline Mg$_3$Sb$_2$~\cite{song2017simultaneous}. The quantitative consistency between theory and experiment reinforces the reliability of the computational framework employed in this work, particularly regarding the treatment of electron–phonon coupling and temperature-dependent $\kappa_{\mathrm{L}}$. Consequently, this validated methodology was applied to GaGe$_2$Te without the introduction of empirical fitting parameters.
	
	Unlike Mg$_3$Sb$_2$, GaGe$_2$Te is a layered vdW semiconductor characterized by a highly anisotropic TE response dominated by cross-plane transport. To accurately capture these directional effects, we evaluated its $\mathrm{ZT}$ performance over a lower temperature range (200--500~K), a regime particularly relevant for layered systems with suppressed $\kappa_{\mathrm{L}}$. As shown in Figure~\ref{PF}, GaGe$_2$Te exhibits remarkably high cross-plane $\mathrm{ZT}$ values for both carrier types. Note that our calculated $\mathrm{ZT}$ values represent only the upper limits achievable under ideal conditions. Specialized experimental synthesis techniques and carrier optimization strategies may be required to approach these values in practice. For $n$-type doping, the cross-plane $\mathrm{ZT}$ reaches a maximum of 2.1 at 500~K with an optimal carrier concentration of $4.7\times10^{18}\,\mathrm{cm}^{-3}$. Notably, $p$-type GaGe$_2$Te achieves a peak cross-plane $\mathrm{ZT}$ of 2.7 at a carrier concentration of $2\times10^{18}\,\mathrm{cm}^{-3}$. These values significantly approaching or exceed those of typical layered TE materials, including $n$-type SnSe (2.8 at 773~K)~\cite{chang20183d}, $p$-type GeTe (2.4 at 600~K)~\cite{li2018low}, and $p$-type BiCuSeO (1.69 at 767 K)~\cite{yin2025ultrahigh}. In contrast, the in-plane TE performance is considerably more modest, with maximum $\mathrm{ZT}$ values limited to 0.34 for $n$-type and 0.05 for $p$-type systems at optimized carrier concentrations. This pronounced anisotropy reflects the fundamental structural characteristics of GaGe$_2$Te: an extremely low cross-plane $\kappa_{\mathrm{L}}$ of $0.57~\mathrm{W\,m^{-1}\,K^{-1}}$ at 300~K, combined with unusually high cross-plane $\mu$ facilitated by intrinsically weak POP scattering. Collectively, these features result in a highly favorable balance of transport parameters along the cross-plane direction.
	
	It is noteworthy that the optimal carrier concentration for maximizing $\mathrm{ZT}$ in GaGe$_2$Te is significantly lower than that of conventional layered TE materials. This trend stems from the material's low valley degeneracy; with fewer degenerate pockets at the band edges, the DOS near the Fermi level is reduced, causing the Fermi level to shift deeper into the conduction or valence band at a fixed carrier concentration. Consequently, a lower carrier concentration is sufficient to achieve the optimal trade-off between $\sigma$ and $S$. Overall, these results position GaGe$_2$Te among the highest-performing layered TE materials predicted to date. The exceptional cross-plane $\mathrm{ZT}$, coupled with intrinsically weak ionic polarization and suppressed POP scattering, underscores the efficacy of targeting materials with low $\epsilon_{\mathrm{ion}}$ as a design paradigm for decoupling electron and phonon transport.

    \section*{Conclusion}\label{conclusion}
	In this work, we propose a novel strategy to decouple electron and phonon transport properties in layered compounds exhibiting strong cross-plane dispersion by minimizing the ionic dielectric constant. Through high-throughput DFT calculations performed on 236 layered compounds, we established a correlation between strong intralayer covalent bonding and high cross-plane mobility, identifying 14 candidates with superior cross-plane power factors. Among these, GaGe$_2$Te emerges as a particularly promising thermoelectric material, as substantiated by state-of-the-art DFT calculations. The van der Waals layered architecture of GaGe$_2$Te induces pronounced anisotropy in both electron and phonon transport. Crucially, the predominantly covalent intralayer bonding—evidenced by high integrated crystal orbital bond index values, small Born effective charges, and a minimal ionic dielectric constant—leads to remarkably weak polar optical phonon scattering. This suppression of the dominant Fr\"{o}hlich interaction facilitates exceptionally long carrier relaxation times and high carrier mobilities, particularly for holes along the cross-plane direction. Concurrently, weak interlayer coupling and strong lattice anharmonicity yield an ultralow cross-plane $\kappa_{\mathrm{L}}$. The synergy between a high power factor and low $\kappa_{\mathrm{L}}$ in GaGe$_2$Te results in outstanding cross-plane $\mathrm{ZT}$ values at 500~K for both $p$ and $n$-type doping. Our findings underscore that layered materials possessing strong covalent bonding characteristics offer a viable pathway to decouple electronic and thermal transport, paving the way for the design of ``phonon-glass electron-crystal'' thermoelectric materials.

	\hspace{0.5cm}
	
	% \begin{suppinfo}
		\noindent $\blacksquare$ \textbf{Supporting information} \\
		
		\noindent The Supporting Information is available free of charge on the ACS Publications website at DOI: \\

		\hspace{0.2cm}

		\noindent \textbf{Notes} \\
		The authors declare no competing financial interest.\\
		
		\hspace{0.5cm}

		\section*{Acknowledgements}
		Z.X., Y.W., and J.H. acknowledge the support received from National Natural Science Foundation of China (Grant No. 12374024) and supported by State Key Laboratory for Advanced Metals and Materials, Grant No. 2025-Z15. Y.Y. acknowledges the support by National Natural Science Foundation of China (grant no. 12304115). M.R. and A.S. acknowledge financial support under the National Recovery and Resilience Plan (NRRP), Mission 4, Component 2, Investment 1.1, Call for tender No. 1409 published on 14.9.2022 by the Italian Ministry of 
		University and Research (MUR), funded by the European Union--NextGenerationEU--Project Title StopHEAT – CUP J53D23016150001- Grant Assignment Decree No. 1381 adopted on 01-09-2023 by the Italian Ministry of Ministry of University and Research (MUR).
		
		\setlength{\bibsep}{0.0cm}
		\bibliographystyle{Wiley-chemistry}
		\bibliography{ref}

\end{sloppypar}
\clearpage

\section*{Entry for the Table of Contents}

\noindent\rule{11cm}{2pt}
\begin{minipage}{11cm}
	\includegraphics[width=11cm]{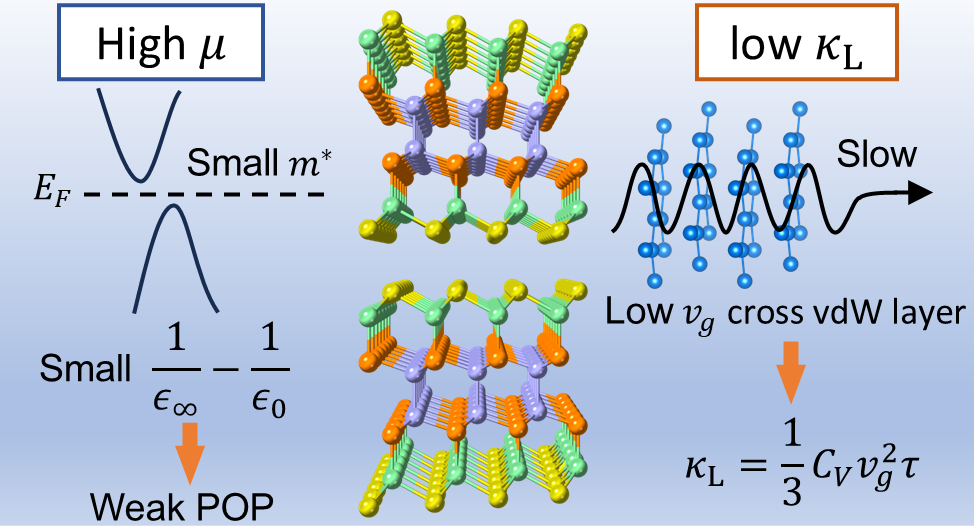}
\end{minipage}
\begin{minipage}{11cm}
	\large\textsf{The small effective mass and high degeneracy caused by intra-chain and inter-chain interactions increase the power factor.}
\end{minipage}
\noindent\rule{11cm}{2pt}

	\end{document}